\documentclass[15pt,a4paper]{article}
\usepackage[ ]{natbib} 
\usepackage[english]{babel}
\usepackage[utf8]{inputenc}
\usepackage[T1]{fontenc}
\usepackage{amsmath,amssymb,amsfonts,amsthm}
\usepackage[fleqn,tbtags]{mathtools}
\usepackage{dsfont} 
\usepackage{authblk}
\usepackage{graphicx}
\usepackage{xcolor}
\usepackage{float}
\usepackage{geometry}
\usepackage{booktabs}
\usepackage{multirow}
\usepackage{xr-hyper} 
\usepackage{hyperref}
\usepackage{array}
\usepackage[labelfont=bf]{caption}
\usepackage{subcaption}
\usepackage{floatpag} 
\floatpagestyle{empty} 
\usepackage{setspace} 
\usepackage{mathpazo} 

\usepackage{parskip} 

\DeclareMathOperator{\Cov}{\mathbb{C}\text{ov}}

\DeclareMathOperator{\influ}{inf}
\DeclareMathOperator{\TPR}{TPR}
\DeclareMathOperator{\FPR}{FPR}
\DeclareMathOperator{\ERR}{ERR}

\DeclareMathOperator{\E}{\mathbb{E}}

\DeclareMathOperator{\sign}{sign}

\newcommand{\norm}[1]{\left\lVert#1\right\rVert}
\newcommand{\transpose}{{}^{\text{\sffamily T}}}
\newcommand{\bigO}{O}

\DeclareMathOperator*{\argmax}{arg\,max}
\DeclareMathOperator*{\argmin}{arg\,min}
\DeclareMathOperator{\asymT}{\text{asym}T}
\DeclareMathOperator{\asymMIP}{\texttt{asymMIP}} 
\DeclareMathOperator{\MIP}{\texttt{MIP}}
\DeclareMathOperator{\asymHIM}{\texttt{asymHIM}}
\DeclareMathOperator{\HIM}{\texttt{HIM}}
\DeclareMathOperator{\rdata}{\texttt{RawData}}
\DeclareMathOperator{\ramm}{\texttt{RaMM}}
\newcommand{\floor}[1]{\lfloor #1 \rfloor} 
\newcommand*\sfref[1]{S\ref{#1}} 

\newtheorem{theorem}{Theorem}

\usepackage[ruled,vlined]{algorithm2e}

\makeatletter
\newcommand*{\addFileDependency}[1]{
  \typeout{(#1)}
  \@addtofilelist{#1}
  \IfFileExists{#1}{}{\typeout{No file #1.}}
}
\makeatother

\newcommand*{\myexternaldocument}[1]{%
    \externaldocument{#1}%
    \addFileDependency{#1.tex}%
    \addFileDependency{#1.aux}%
}

\myexternaldocument{8supplementary}

\title{An algorithm-based multiple detection influence measure for high dimensional regression using expectile}

\author[1,2]{Amadou Barry\footnote{Corresponding author:amadou.barry@mail.mcgill.ca} }
\author[3]{Nikhil Bhagwat}
\author[3]{Bratislav Misic}
\author[3,4,5]{Jean-Baptiste Poline}
\author[1,2,5,6]{Celia M. T. Greenwood}

\affil[1]{Lady Davis Institute, Jewish General Hospital, Montréal, Québec, Canada}
\affil[2]{Department of Epidemiology, Biostatistics and Occupational Health,
McGill University, Montréal, Québec, Canada}
\affil[3]{Faculty of Medicine, Department of Neurology and Neurosurgery, Montreal Neurological Institute and Hospital, McConnell Brain Imaging Centre, McGill University, Montréal, Québec, Canada}
\affil[4]{Henry H. Wheeler Jr. Brain Imaging Center, Helen Wills Neuroscience Institute, University of California, Berkeley, CA, United States}
\affil[5]{Ludmer Centre for Neuroinformatics \& Mental Health, McGill University, Montreal, QC, Canada}
\affil[6]{Department of Oncology and Human Genetics, McGill University, Montréal, Québec, Canada}

\date{\today}

\begin{document}
\maketitle

\begin{abstract}
~~\\
The identification of influential observations is an important part of data analysis that can prevent erroneous conclusions drawn from biased estimators.
However, in high dimensional data, this identification is challenging. Classical and recently-developed methods often perform poorly when there are multiple influential observations in the same dataset. In particular, current methods can fail when there is \emph{masking:} several influential observations with similar characteristics,  or \emph{swamping:} when the influential observations are near the boundary of the space spanned by well-behaved observations. Therefore, we propose an algorithm-based, multi-step, multiple detection procedure to identify influential observations that addresses current limitations. Our three-step algorithm to identify and capture undesirable variability in the data, $\asymMIP,$ is based on two complementary statistics, inspired by asymmetric correlations, and built on expectiles. Simulations demonstrate higher detection power than competing methods. Use of the resulting asymptotic distribution leads to detection of influential observations without the need for computationally demanding procedures such as the bootstrap. The application of our method to the Autism Brain Imaging Data Exchange neuroimaging dataset resulted in a more balanced and accurate prediction of brain maturity based on cortical thickness. See our GitHub for a free R package that implements our algorithm: \texttt{asymMIP}  (\url{github.com/AmBarry/hidetify}).

\end{abstract}


\section{Introduction}\label{Intro}

In data science, signals of interest are often mixed with different type of error or noise, e.g., variables measured with error or corrupted.  Furthermore, the increase in the number of predictors $(p)$ increases the risk of having at least one error in the data, particularly when $p>>n$, where $n$ is the number of observations, or when data contain measurements obtained from multiple sources. For example, in neuroimaging, structural and functional magnetic resonance imaging (fMRI) scans are often plagued by errors occurring during acquisition (head motion, low quality images due to poorly calibrated MRI machines), or the essential preprocessing steps (alignment, coregistration and normalization). In genomics, gene expression microarray data and high-throughput sequencing errors may arise due to degraded samples or polymerase chain reaction amplification bias. Such corrupted observations may negatively impact inference and result in biased estimators; if this is the case they are usually termed \emph{influential} observations. Note that a corrupted observation is not necessarily influential. 

In classical statistical setting \((n>p),\) a panoply of robust statistical methods have been proposed to mitigate the presence of  influential observations \citep{chatterjee_influential_1986, Rousseeuw1987}. Today, many researchers are actively working on extending these robust methods to high-dimensional data, with more or less success. Some have proposed the adaptation of robust methods \citep{NoureddineElKaroui2013, PoLingLoh2017, Prasad2020} less prone to influential observations, others have explored the development of diagnostic methods and identification of influential observations in the regression framework \citep{she_outlier_2010, jeong_effect_2018, zhao_high_dimensional_2013, asymHimBarry2019} and the multivariate framework \citep{filzmoser_outlier_2008, fritsch_detecting_2012, ro_outlier_2015}. Certainly, the study and identification of high dimensional influential observations is known to be challenging. Not only robust estimators are hard to compute in high dimensions \citep{Diakonikolas2021}, but also traditional diagnostic tools fail in this context{\textemdash}for example the gram matrix \((\boldsymbol{X}\transpose\boldsymbol{X}),\) is not invertible. On top of that, it is difficult to visualize high dimensional data whose dimensionality and complexity increase the occurrence probability of contaminated observations.


It is in this line of research of devising  robust methods that are less prone to influential observations that, \citet{asymHimBarry2019} introduced an expectile based asymmetric correlation influence measure for high dimensional linear regression ($\asymHIM$ : asymmetric high dimensional influence measure). They showed that the $\asymHIM$ influence measure generalizes and outperforms a leave-one-out symmetric influence measure \citep{zhao_high_dimensional_2013} built on the sure independence screening concept of \citet{fan_sure_2008}. In fact, \citet{zhao_high_dimensional_2013}'s influence measure can be considered to be an $\asymHIM$ influence measure with expectile \(\tau=0.5.\) Another proposal by \citet{wang_outlier_2017} identifies influential observations through marginal correlations, however this approach requires computationally intensive bootstrapping to identify an appropriate threshold for calling the outliers. 

\begin{figure}
\centering
\includegraphics[width=\textwidth]{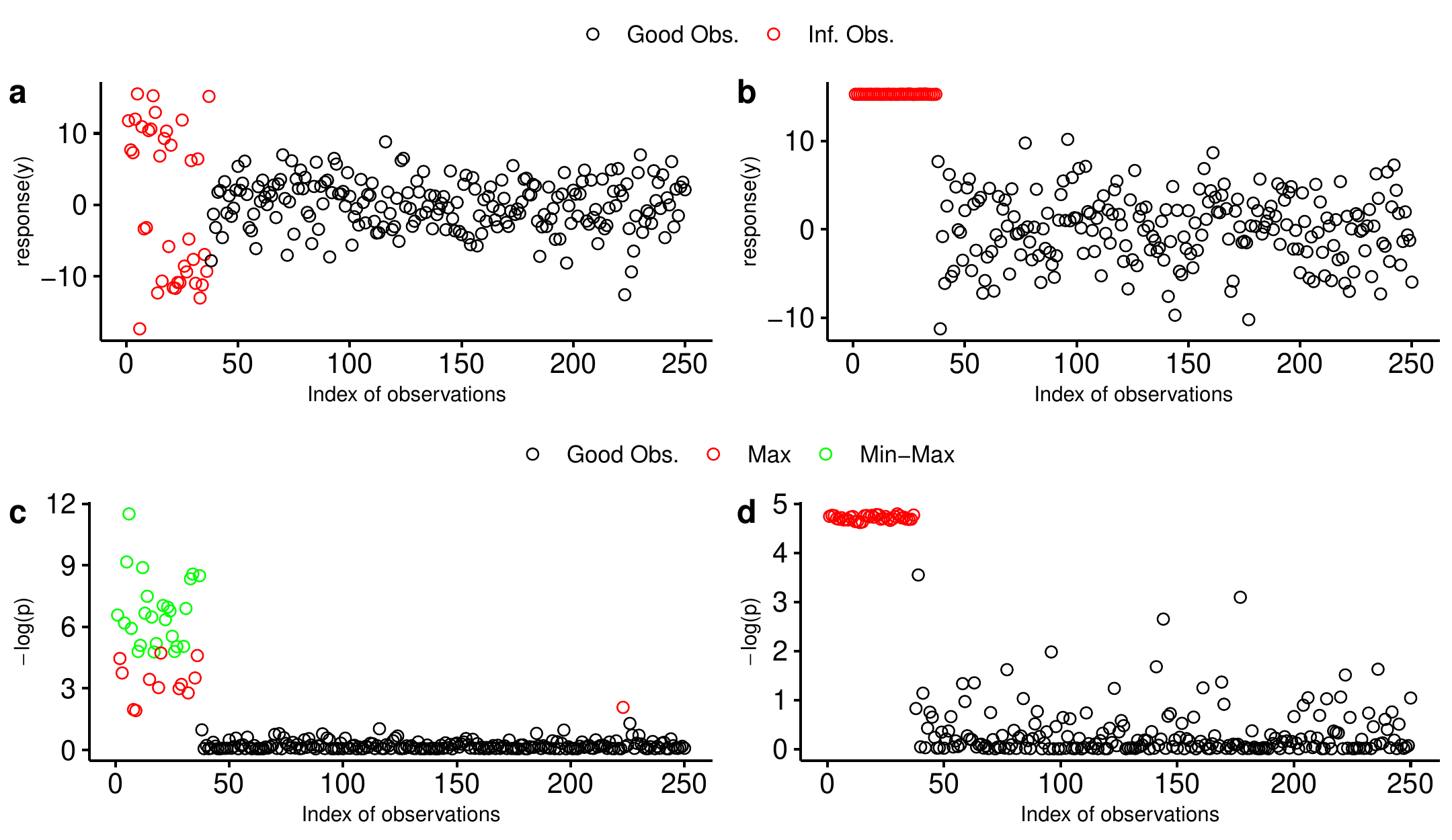}
\caption{Illustration of the swamping (Figure \ref{fig:plot_ilustration}a) and masking (Figure \ref{fig:plot_ilustration}b) effects and application of the $\asymT_{\min,k,m}$ (Figure \ref{fig:plot_ilustration}c) and $\asymT_{\max,k,m}$ (Figure \ref{fig:plot_ilustration}d) influence measures. Figure \ref{fig:plot_ilustration}a and \ref{fig:plot_ilustration}b show the value of the response variable (y) on the y-axis. Figure \ref{fig:plot_ilustration}c and \ref{fig:plot_ilustration}d show the log p-value (-log(p)) of the observations affected by a swamping and masking effects, respectively. The labels Good Obs. and Inf. Obs. indicate good and influential observations, respectively. The Max and Min-Max labels represent the observations identified as influential by the $\asymT_{\max,k,m}$ measure only and by the $\asymT_{\min,k,m}$ and $\asymT_{\max,k,m}$ measures, respectively. The $\asymMIP$ measure identifies $100\%$ of the influential observations in either case without any false positive.}\label{fig:plot_ilustration}%
\end{figure}

However, these methods use a \textit{single detection technique} assuming contamination by only one influential observation. Not only is this assumption unlikely to be true for large, complex real data, but also it has been shown that when this assumption is violated, single detection technique-based influence measures will perform poorly. In particular,  \textit{masking}, where an  influential observation is masked or hidden by other similar influential observations,  or \textit{swamping}, where a non-influential or good observation is falsely classified as influential due to nearby influential observations, often confuse single detection-based measures of influence \citep{nurunnabi_procedures_2014, Zhao_2019, asymHimBarry2019}. The effects of masking and swamping on the response variable are illustrated in Figure \ref{fig:plot_ilustration}a and \ref{fig:plot_ilustration}b, with data generated from the simulation models in Section \ref{Sim}. Of course, these effects also influence the predictor space, but are more difficult to visualize there.  

In response to this challenge, recent propositions \citep{wang_multiple_2018, wang_multiple_case_2018, Zhao_2019} implement a variant of multiple case deletion or group deletion \citep{belsley_regression_1980} originally developed in the classical statistical framework. These methods try to find a clean subset of the initial sample, and then apply  single detection techniques to test for influence of each observation left out. 

Unfortunately, all these algorithm-based influence measures, except \citet{Zhao_2019}'s method, rely on suggestive approaches and computationally expensive methods such as bootstrap to determine the influential nature of the observations. \citet{Zhao_2019}'s algorithm-based influence measure called MIP (multiple influential point detection) is computationally efficient and it outperforms well-known high dimensional robust statistical methods \citep{WangLiJiang2007, Maronna2011, smuclerRobustSparseEstimators2017}. However, the MIP procedure has a low power to detect influential observations, potentially due to a poor asymptotic distribution. Furthermore, \citet{wang_multiple_case_2018} showed that the MIP measure is ineffective at overcoming the swamping effect and that its power is compromised due to the very low false positive rate. In contrast to \citet{zhao_high_dimensional_2013}'s $\HIM$ measure, the $\asymHIM$ influence measure is powerful at identifying influential observations in high dimension.


The $\asymHIM$ influence measure \citep{asymHimBarry2019} is powerful at identifying influential observations in high dimensional data, but performance deteriorates when there is swamping or masking. Therefore, in this paper, we propose a three step algorithm-based group deletion procedure to identify influential observations in high dimensional regression, that we call $\asymMIP$
(asymmetric multiple influential point detection). The first two steps are built on new statistics, $\text{asym}T_{\min,k,m} \mbox{ and } \ \text{asym}T_{\max,k,m}$, to mitigate swamping and masking effects, respectively. The last step validates observations identified as influential in the previous steps through the application of our single detection based influence measure, $\asymHIM.$ Through the availability of an asymptotic distribution for $\asymMIP$ when there are no influential observations (a null distribution), this method is computationally efficient.

The strengths and weaknesses of the new influence measures $(\text{asym}T_{\min,k,m} \mbox{ and } \ \text{asym}T_{\max,k,m})$ are illustrated in Figure \ref{fig:plot_ilustration}c and \ref{fig:plot_ilustration}d. It is remarkable that \(\text{asym}T_{\max,k,m}\) captures  all the influential observations in the presence of masking, while \(\text{asym}T_{\min,k,m}\) identifies the majority of the influential observations in the presence of swamping. However, the aggressive \(\text{asym}T_{\max,k,m}\) finds some false positives in the presence of swamping, while the conservative \(\text{asym}T_{\min,k,m}\) has low power in the presence of masking. Nevertheless, our combined three-step $\asymMIP$ influence measure identifies all the influential observations without any false positives in either case. As shown in the Simulation section, use of this $\asymMIP$ measure will result, in subsequent analyses, in a sparse model with less biased and more accurate estimators.

In the next section, Section \ref{MM}, we introduce the concept of expectile and briefly review the $\asymHIM$ measure of \citet{asymHimBarry2019}. Next, we present the new influence measures along with the proposed algorithm to mitigate the masking and swamping effects. Section \ref{MM} concludes with the derivation of the asymptotic properties of the two influence measures. In Section \ref{Sim} we evaluate the performance of our algorithm with simulations, and in Section \ref{app}, we use our $\asymMIP$ influence measure to quality check the Autism Brain Imaging Data Exchange (ABIDE) neuroimaging dataset before predicting the brain maturity of the control group based on 299,569 measures of cortical thickness.

\section{Models and Methods} \label{MM}
In this section we introduce the expectile statistics and we define the asymmetric influence measure. 
Then, we present the multiple detection algorithm and we state the theorem on the asymptotic distribution of the statistics involved in the algorithm. We start with some notation.

\textbf{Notation.} We write the vectors \((\boldsymbol{x}\in\mathbb{R}^p)\) in lower bold letters and we represent the matrices \((\boldsymbol{X}\in\mathbb{R}^{n\times p)}\) in capital bold letters. The symbol \(\ \mathbb{I}_{p\times p}\ \) or \(\ \mathbb{I}_{p}\) represents the identity matrix and is noted as \(\mathbb{I}\) when the dimension is implicitly known. The symbol \(\mathds{1}(t<0)\) is the indicator function and is equal to 1 if \(t<0\) and to 0 otherwise. We denote the symbol \(\norm{\cdot}\) as the \(l_2\) norm and  we define the number of elements in the set \(S\) as \(\lvert S \rvert.\) We denote by \(S_{inf}\) the set of influential observations, and \(n_{inf}=\lvert S_{inf} \rvert\) its cardinality. We denote by \(\widehat{S}_{inf}\) the estimator of the set \(S_{inf}\) and \(S_{inf}^{c}\) its complement. We define  \(\text{Supp}(\boldsymbol{\beta}) = \lbrace j \ | \ \beta_j \ne 0, \mbox{ for } j = 1,\ \dots,\ p \rbrace\) as the set of regressor parameters with non-zero effect size (active set) and by \(\text{Supp}^c(\boldsymbol{\beta}) = \lbrace 1,\ \dots,\ p \rbrace \setminus \text{Supp}(\boldsymbol{\beta})\) its complement.

\subsection{Expectile}

Expectile statistics are asymmetric weighted averages that characterize the cumulative distribution function (c.d.f) of a random variable (\citet{GEEE_Barry2018}). Expectiles summarize the c.d.f while accounting for the underlying heterogeneity. In this regard, expectiles play a similar statistical role to quantile statistics, except that quantiles are order statistics while expectiles are weighted averages, and this interpretation difference is accompanied by significant computational advantages. Further introduction and details about expectiles and quantiles can be found in (\citet{newey_asymmetric_1987}).

The expectile, \(\mu_{\tau}(y),\) of level \(\tau\in(0,\ 1)\) of a continuous random variable (r.v.) is defined as:
\begin{equation}\label{exp1_geee}
\mu_{\tau}(y) = \argmin_{\theta\in\mathbb{R}} \E\{\mathcal{R}_\tau(y-\theta)\},
\end{equation}
where the function \(\mathcal{R}_\tau(t)=\lvert \tau-\mathds{1}(t\leq 0)\rvert \cdot t^2\) is an asymmetric square loss function that assigns weights \(\tau\) and \(1-\tau\) to positive and negative deviations, respectively.

Equation (\ref{exp1_geee}) leads to a definition of the expectile of level \(\tau\) through the following functional equation:
\begin{equation*}
   \mu_{\tau}= \E\Bigg[ \frac{\psi_{\tau}(y-\mu_{\tau})}{\E\big[\psi_{\tau}(y-\mu_{\tau})\big]}y\Bigg],
\end{equation*}
where \(\psi_{\tau}(t)=\lvert \tau-\mathds{1}(t\leq 0)\rvert\) is the piecewise linear function. The expectile of level \(\tau=0.5\) corresponds to the well-known expectation \(\ \E[y] = \mu = \mu_{0.5}.\) Given a random sample, \(\lbrace(y_i)\rbrace_{i=1}^{n},\) the \(\tau\)th empirical expectile is defined as:\\
\begin{equation*}
    \widehat{\mu}_{\tau}= \argmin_{\theta\in\mathbb{R}} \frac{1}{n}\sum_{i=1}^{n}\mathcal{R}_\tau(y_i-\theta) = \sum_{i=1}^{n}\frac{\psi_{\tau}(y_i-\widehat{\mu}_{\tau})}{\sum_{i=1}^{n}\psi_{\tau}(y_i-\widehat{\mu}_{\tau})}y_i.
\end{equation*}
The empirical expectile can be computed using the iterative reweighted least square (IRLS) algorithm. Since the asymmetric square loss function is convex and continuously differentiable, IRLS converges very fast. The expectreg package (\citet{expectreg}, \citet{SobotkaKneib2012}, \citet{SchnabelEilers2009}) implemented in \citet{rcran} can be used to generate expectiles.

\subsection{Asymmetric influence measure} \label{asymHim}

\citet{asymHimBarry2019} proposed an asymmetric high dimensional influence measure (asymHIM) based on marginal asymmetric correlations. The asymmetric correlation is a function of asymmetric variances/covariances, centered around the chosen expectile of the random variable distribution. This asymmetric variance{\textemdash}defined at different expectiles{\textemdash}will deferentially capture the variability, and this may be helpful in the presence of skewed or heavy-tailed distributions. 

For a fixed \(\tau\in(0,\ 1),\) the asymmetric covariance and the asymmetric correlation between two random variables \(x\) and \(y\) are defined, respectively, as:

\begin{equation*}
\Cov_{\tau}(x,y) = \E\Big[(x -\mu_{\tau}(x))(y -\mu_{\tau}(y)) \Big],  
    \quad \rho_{\tau}(x,y) = \frac{\Cov_{\tau}(x,y)}  {\sigma_{\tau}(x)\sigma_{\tau}(y)}, 
\end{equation*}

where \(\sigma_{\tau}^{2}(y) = \E\Big[(y-\mu_{\tau}(y))(y-\mu_{\tau}(y)\Big]\) is the asymmetric variance and \(\sigma_{\tau}^2(x)\) is defined similarly. The standard covariance and correlation definitions apply when $\tau = 0.5$.

Similarly, given two empirical vectors, \((\boldsymbol{x}\transpose,\ \boldsymbol{y}\transpose)\transpose = (x_{1},\ \ldots, \ x_{n}, \ y_{1},\ \ldots, \ y_{n} )\transpose\) of size \(n,\) the corresponding sample estimate of the asymmetric covariance and correlation function, for a fixed \(\tau\in(0,1),\) can be defined as:

\begin{equation*} 
\widehat{\Cov}_{\tau}(\boldsymbol{x},\boldsymbol{y}) = n^{-1}\sum_{i=1}(x_{i}-\widehat{\mu}_{\tau}(\boldsymbol{x}))(y_{i}-\widehat{\mu}_{\tau}(\boldsymbol{y})), \quad
\widehat{\rho}_{\tau}(\boldsymbol{x},\boldsymbol{y})= \frac{\widehat{\Cov}_{\tau}(\boldsymbol{x},\boldsymbol{y})}  {\widehat{\sigma}_{\tau}(\boldsymbol{x})\widehat{\sigma}_{\tau}(\boldsymbol{y})},
\end{equation*}

where \( \ \widehat{\sigma}_{\tau}^2(\boldsymbol{x})=n^{-1}\sum_{i=1}(x_{i}-\widehat{\mu}_{\tau}(x_{i}))^2 \ \) is the empirical asymmetric variance and \( \ \widehat{\sigma}_{\tau}^2(\boldsymbol{y})\ \) is defined similarly. As an illustration of the usefulness of the asymmetric covariance, we have generated random variables with three different distributions: standard normal \((\mathcal{N}(0,1)),\ \) Student \((\mathcal{T}_3)\ \) with 3 degrees of freedom, and Chi-square \((\chi_{2}(3)).\) Then, we have estimated the asymmetric covariances at 3 different expectiles, \(\tau\in(0.25,\ 0.5,\ 0.75).\) The results, displayed in \textbf{Table \ref{tab:tab1}}, show that the asymmetric variance captures the heaviness of the tails. For example, we see larger asymmetric variances for the left tail than the right for the asymmetric Chi-squared distribution. 

\begin{table}[h!]
\caption{Empirical asymmetric variance of 3 random variables (Normal, Student and Chi-square) centered at 3 different expectiles, $\tau\in(0.25,\ 0.5,\ 0.75).$ } \label{tab:tab1} 
\centering
\begin{tabular}{l*{3}{c}}
\toprule
 & \multicolumn{3}{c}{$\tau$ } \\
 \cmidrule(l){2-4} 
 & 0.25 & 0.5 & 0.75 \\
 \cmidrule(l){2-4}
 $\mathcal{N}(0,1)$ & 1.29 & 0.92 & 1.27  \\
 $\mathcal{T}_3$ & 2.73 & 2.23 & 2.77  \\
  $\chi_{2}(3)$ & 9.81 & 6.83 & 7.72  \\
\bottomrule
\end{tabular}%
\end{table}

Next, we review the asymmetric influence measure (asymHIM) proposed by \citet{asymHimBarry2019}. We assume that the non-influential observations in the data are independently and identically distributed and are generated by the following regression model:

\begin{equation}\label{eq:lm} 
y = \beta_0 + \boldsymbol{x}\transpose\boldsymbol{\beta}_1 +  \varepsilon, 
\end{equation}

where \(y\) is the response variable, \(\boldsymbol{x}\) is the associated p-vector of predictors, and \(\boldsymbol{\beta}=(\beta_0,\ \boldsymbol{\beta}_1\transpose)\transpose\) the parameter vector of interest. The random error \(\varepsilon \sim \mathcal{N}(0,\ 1)\) is assumed to follow a standard normal distribution. To be specific, samples generated by model equation (\ref{eq:lm}) are clean (not contaminated) samples.

Given a random sample, \(\ [\boldsymbol{X};\ \boldsymbol{y}]=[(\boldsymbol{x}_1,\ \ldots, \ \boldsymbol{x}_p);\ (y_1,\ \ldots,\ y_n)\transpose],\ \) generated by model equation (\ref{eq:lm}), where \(\ \boldsymbol{x}_j=(x_{1j},\ \ldots,\ x_{nj})\transpose,\ \mbox{ for } \ 1\leq j \leq p \
\mbox{ and } \ p > n,\) \citet{asymHimBarry2019} employed a leave-one-out technique to quantify the influence of the \(k\)th observation using the following measure:

\begin{equation} \label{eq:single_him} 
\mathcal{D}_{\tau k}= \frac{1}{p}\sum_{j=1}^{p} \Big(\widehat{\rho}_{\tau j} - \widehat{\rho}_{\tau j}^{(k)}\Big)^2,
\end{equation}

built on the asymmetric empirical correlation, where

\begin{equation*}
\widehat{\rho}_{\tau j}^{(k)}=\frac{\sum_{i=1, i\ne k}^{n}(x_{ij}-\widehat{\mu}_{\tau}^{(k)}(x_{ij}))(y_i-\widehat{\mu}_{\tau}^{(k)}(y_i))}
{(n-1)\widehat{\sigma}_{\tau}^{(k)}(x_{ij})\widehat{\sigma}_{\tau}^{(k)}(y_i)}, \; j=1,\ \ldots, \ p; \ k=1,\ \ldots,\ n.
\end{equation*}


The quantities with superscript $k$ indicate estimates with the $k^{th}$ observation removed. This measure in equation (\ref{eq:single_him}) with \(\tau=0.5\) corresponds to \citet{zhao_high_dimensional_2013}'s high dimensional influence measure (HIM). However, the addition of asymmetric expectiles improves sensitivity for finding outlier observations.

The asymmetric influence measure \(\mathcal{D}_{\tau k}\) is a function of the asymmetric point \(\tau\in(0,\ 1),\) and hence different influential point measures can be derived by varying $\tau$. After exploring various scenarios with simulations, \citet{asymHimBarry2019} 
proposed using the sum of  \(\mathcal{D}_{\tau k}\) for several values of $\tau$ 
to quantify the influence of the \(k\)-th observation. Therefore, the asymHIM measure for a sequence of asymmetric points \((\tau_1,\ \ldots,\ \tau_q),\) was defined as:
\begin{equation}\label{eq:wahdim)} 
\textit{asym}{\mathcal{D}}_{k}= \frac{1}{p}\sum_{l=1}^{q}\sum_{j=1}^{p} \Big(\widehat{\rho}_{\tau_{l} j} - 
\widehat{\rho}_{\tau_{l} j}^{(k)}\Big)^2 = \sum_{l=1}^{q}\mathcal{D}_{\tau_{l} k}, \quad k = 1, \ \ldots,\ n,
\end{equation}
In their previous work, the statistical distribution of this statistic was derived, and \citet{asymHimBarry2019} suggested using the asymHIM measure with the three expectiles \(\tau\in (0.25,\ 0.5,\ 0.75)\ \).

Despite its attractive properties, the asymHIM measure does not always achieve desired performance. Indeed, the null distribution of the asymHIM measure is established under the assumption of an uncontaminated or cleaned sample, usually a untenable and unverifiable condition, since most of the time large, high dimensional datasets are likely to contain some contaminated data. Furthermore, data patterns known as masking and swamping adversely affect performance of a single detection-based technique. 



\subsection{Multiple influential points detection procedure} \label{asymMip}

For identification of influential observations, the challenges arising from the dual phenomena of masking and swamping seem, at first glance, difficult to overcome. For example, to protect against masking, one might look for a more aggressive influence measure, but this may result in a high rate of falsely identified outliers. In contrast, a conservative approach might be selected to protect against swamping which may have many false negatives.
\citet{asymHimBarry2019} showed that the asymmetric influence measure based on the maximum of a sequence of expectiles for different values of $\tau$ was an aggressive measure identifying too many influential points, whereas the minimum of this same sequence was too conservative in identifying the influential observations.

Therefore, in the classical linear regression literature (\citet{belsley_regression_1980}, \citet{nurunnabi_procedures_2014}, \citet{roberts_adaptive_2015}), solutions such as the leave-many-out or group deletion procedure have been proposed. The central idea revolves around finding a non-contaminated sub-sample (a sample without any influential observations), and then using this subsample as a reference set against which the influence of the remaining observations is tested. Given that the number of influential observations is unknown, several subsets must be sampled in order to increase the likelihood of finding a clean subsample. This concept has been recently adapted to high dimensional influence measures (\citet{wang_multiple_2018}, \citet{wang_multiple_case_2018}, \citet{Zhao_2019}).

In this paper, we use a min-max random deletion algorithm, along other asymmetric influence measures, to mitigate the detrimental effects of swamping and masking on asymHIM. Before presenting the algorithm, we introduce some additional notation.

Let \(n_{inf} = \lvert S_{inf} \rvert\) be the number of influential observations and \(n - n_{inf} = \lvert S_{inf}^c \rvert\) the number of indices of the non influential observations. For any observation \(k,\) denote by \(A_1, \cdots, A_m \subset \lbrace 1, \cdots, n \rbrace \setminus \lbrace k \rbrace\) the subsets of size \(n_k = \lvert A_r \rvert,\) uniformly and randomly drawn with replacement, and \(A_r^{(+k)}= A_r \cup \lbrace k \rbrace\) the subset of size \(n_{sub}=n_k +1,\) for \(r\in\lbrace 1,\cdots,m\rbrace.\) Additionally, for \(1 \leq r \leq m,\) let \(B_r\) be the subset of the non-influential observation's indices, in \(A_r,\) and \(N_{B_r} = \lvert Br \rvert\) its size. Likewise, let \(O_r=A_r\setminus B_r\) be the subset of the influential observation's indices in \(A_r.\)

With this notation in hand, we define a new asymmetric influence measure for the \(k\)-th observation in each subset \(A_r, r=1,\cdots, m\), for a fixed \(\tau,\) as:
\begin{equation} \label{eq:single_him_subset)} 
\mathcal{D}_{\tau r k}= p^{-1} \norm{ \widehat{\rho}_{\tau A_{r}^{(+k)} } - \widehat{\rho}_{\tau A_{r} } }^2, \enspace \mbox{ and } \enspace
\text{asym}{\mathcal{D}}_{rk} = \sum_{l=1}^{q}\mathcal{D}_{\tau_{l} r k},
\end{equation}
where \(\widehat{\rho}_{\tau A_{r}^{(+k)} }\) and \(\widehat{\rho}_{\tau A_{r} }\) are respectively the estimated asymmetric correlation based on \(A_{r}^{(+k)}\) and \(A_{r}.\)

For each random subset, \(A_r, r=1,\cdots,m,\) the contribution of the \(k\)-th observation is summarized by a sequence of asymmetric influence measures selected from \(\tau\in(0,1)\). In \citet{asymHimBarry2019}, we observed that using only the minimum of this sequence was too conservative, while using the maximum was too aggressive. Therefore, here we propose to implement a random min-max multiple deletion ($rmmm$) algorithm using the following asymmetric influence measures:
\begin{equation} \label{eq:min_max)} 
\text{asym}T_{\min,k,m} = \min_{1\leq r\leq m} \min_{1\leq l\leq q} n_{sub}^2 \mathcal{D}_{\tau_{l} r k}, \enspace \mbox{ and } \enspace
\text{asym}T_{\max,k,m} = \max_{1\leq r\leq m} n_{sub}^2 \text{asym}{\mathcal{D}}_{rk}.
\end{equation}
Subsets are chosen randomly and uniformly with replacement. Hence, as mentioned by \citet{Zhao_2019}, the asymmetric marginal correlation based on these subsets can be seen as a kind of perturbation to the marginal correlations based on the whole sample. Both statistics, \(\text{asym}T_{\min,k,m}\) and \(\text{asym}T_{\max,k,m},\) measure the distance of the influence measures to the data centre based on the randomly sampled datasets.

Since the asymmetric influence measure \(\text{asym}T_{\min,k,m}\) includes the minimum influence measure proposed by \citet{zhao_high_dimensional_2013}, it can be expected to be  more effective in correcting for swamping. Furthermore \citet{asymHimBarry2019} showed that the asymmetric influence measure \(\text{asym}{\mathcal{D}}_{k}\) is more efficient than the HIM measure proposed by \citet{zhao_high_dimensional_2013}, so our asymmetric influence measure \(\text{asym}T_{\max,k,m}\) here can also be expected to perform better in the presence of masking. 
The details of the algorithm are summarized in Algorithm \ref{rmmmd_algo}. The parameter \(m\) is the number of drawn sub-samples and \(n_k\) their size. There is one additional parameter, \(\omega\),
which controls the influential set size in the Min Step procedure to ensure that the following step does not end up with a very small sample size. 

\begin{algorithm}[H]
\DontPrintSemicolon
\SetAlgoLined
\SetKwProg{Proc}{Procedure}{}{}

\KwIn{$S_{\text{total}}=\lbrace 1,\ \ldots,\ n\rbrace, \ n_k,\ m, \ $ and $\omega$}

\Proc{Min-Max}{
 \While{$\lvert \widehat{S}^{c}_{\max,\ \influ}\rvert \leq n/2$}{
 \Proc{Min STEP - Mitigate Swamping}{
    \ForEach{ each $k\in S_{\text{total}}\;$ }{
        Generate $m$ subsets of size $n_k$ from $S_{\text{total}}\setminus \lbrace k \rbrace$ \;
        Compute $\text{asym}T_{\min,\ k,\ m}$\;
    }
    Estimate $\widehat{S}_{\min,\ \influ} = \lbrace k_1, \ldots, k_{n_k}| P(\chi_2^1>\text{asym}T_{\min,k_1,m})< \ldots < P(\chi_2^1>\text{asym}T_{\min,k_{n_k},m}) <\alpha/n_k, 1\leq k \leq n\rbrace,$
    where $\lvert\widehat{S}_{\min,\influ}\rvert = \min(\omega n, n_{\min, \influ}), \omega\in(0,1)$ and  $n_{\min, \influ}$ is the number of influential observations acquired from the Min STEP\; 
    Update $S_{\text{total}} \gets S_{\text{total}} \setminus \widehat{S}_{\min,\influ}$ \; 
 }
 \Proc{Max STEP - Mitigate Masking}{
    \ForEach{ each $k\in S_{\text{total}}$ }{
        Generate $m$ subsets of size $n_k$ from $S_{\text{total}}\setminus \lbrace k \rbrace$ \;
        Compute $\text{asym}T_{\max,k,m}$   \;
    }
        Estimate $\widehat{S}_{\max,\influ}$ \;
        Update $S_{\text{total}} \gets S_{\text{total}} \setminus \widehat{S}_{\max,\influ}$ \; 
 }
}
\Proc{Validation STEP}{
\ForEach{ each $k\in \widehat{S}_{\max,\influ}$ }{
  Compute $\textit{asym}{\mathcal{D}}_{k}$ using the updated sample $S_{\text{total}}$ as clean sample  \;
  Estimate $\widehat{S}_{\influ}$ \;
}
}
return $\widehat{S}_{\influ}$  and $\widehat{S}^c_{\influ}$ \;
}
\caption{\texttt{RaMM} algorithm}\label{rmmmd_algo}
\end{algorithm}

In summary, our algorithm consists of three main steps. 
The first two steps are iterated until convergence. The initial step (Min Step) applies the \(\text{asym}T_{\min,k_{n_k},m}\) influence measure to protect against swamping, and eliminates influential observations with a large or moderate effect. In this step, the number of excluded observations is controlled by the parameter \(\omega,\) to ensure that the next step contains sufficient samples. In the second stage (Max Step) of the iterative cycle, the newly defined uncontaminated sample is used to calculate the aggressive \(\text{asym}T_{\max,k,m}\) influence measure, which protects against masking. If the clean set sample size is small \((\leq n/2)\) then the iterative loop continues to run to identify other influential observations with a large impact. In practice, Algorithm \ref{rmmmd_algo} is computationally efficient and usually identifies influential observations after one iteration. Finally, testing is applied in the last step (Validation Step) to confirm the status of those observations deemed influential. The algorithm's sensitivity to the different parameters is evaluated in the simulation section, Section \ref{Sim}.

We complete this section by examining the asymptotic behaviour of the asymmetric influence statistics \(\text{asym}T_{\min,k,m}\) and \(\text{asym}T_{\max,k,m},\). This asymptotic distribution is used to define a statistical cutoff threshold under the null hypothesis and at the same time enhance the computational effectiveness of \textbf{Algorithm} \ref{rmmmd_algo}. Asymptotic results are established under the following conditions.

\textbf{C1}. For any fixed \(\tau \in (0,1)\) and \(1 \leq j \leq p, \; \rho_{\tau j}\) is constant and does not change as \(p\) increases.

\textbf{C2}. The asymmetric covariance matrix \(\boldsymbol{\Sigma}_{\tau} = \Cov_{\tau}(\boldsymbol{x}) = \E[(\boldsymbol{x} - \mu_{\tau}(\boldsymbol{x})) (\boldsymbol{x} - \mu_{\tau}(\boldsymbol{x}))\transpose],\) with the eigen decomposition \(\boldsymbol{\Sigma}_{\tau} = \sum_{j=1}^{p} \lambda_{\tau j}\boldsymbol{u}_j\boldsymbol{u}_j\transpose,\) is assumed to satisfy \(l_{p\tau} = \sum_{j=1}^{p}\lambda_{\tau j}^2 = \bigO(p^r)\) for some \(0\leq r < 2\) and for any fixed \(\tau \in (0,1).\)

\textbf{C3}. The predictor \(\boldsymbol{x}_i\transpose\) follows a multivariate normal distribution and the random noise \(\varepsilon_i\) follows a standard normal distribution.

\textbf{C4}. Let \((Q_{\tau},R_{\tau}) =\Big((\widehat{\mu}_{{\tau}}-\mu_{{\tau}})/\sigma_{{\tau}}, \sigma_{{\tau}}/\widehat{\sigma}_{{\tau}}-1\Big), \quad S_{Q_{y\tau}} = \limsup\limits_{n\rightarrow \infty}\E[n^{1/2}Q_{y\tau}]^8, \mbox{ and } S_{R_{y\tau}} = \limsup\limits_{n\rightarrow\infty}\E[n^{1/2}R_{y\tau}]^8.\) Assume that \(S_{Q_{y\tau}}, \; \mbox{ and } S_{R_{y\tau}}\) are finite. Furthermore, assume there exist constants \(0<K,C<\infty,\) independent of \(n,p, \mbox{ and } \tau,\) such that for any \(t>0,\)
\begin{align*} 
\max_{j\leq p} \mathbb{P} \Big(\lvert \mu_{{\tau j}}-\widehat{\mu}_{{\tau j}} \rvert > t/\sqrt{n}\Big) \leq C \exp(-t^{2}/K) \\ 
\max_{j\leq p} \mathbb{P} \Big(\lvert \widehat{\sigma}_{{\tau j}}/\sigma_{{\tau j}} -1 \rvert > t/\sqrt{n}\Big) \leq C \exp\Big(-\min(t/K,t^{2}/K^{2})\Big). \\ 
\end{align*}
The stated conditions are similar to those stated by \citet{Zhao_2019}. Condition \textbf{C1} assumes an unknown fixed correlation \(\rho_{\tau j}\) between the response and the predictor \(j\) for any fixed \(\tau\) and independently of \(p.\) Condition \textbf{C2} permits high values of the eigenvalues of the covariance matrix \(\boldsymbol{\Sigma}_{\tau},\) but at a rate controlled by the dimensionality \(p.\) A sufficient condition would be that \(\max_{1\leq j \leq p} \lambda_{\tau j}\) be bounded. The normality assumption, Condition \textbf{C3}, is used, among others, to ensures independence across columns of the matrix \(\boldsymbol{X}.\) For reading convenience we introduce the following notation: \(\mu_{j\tau}=\mu_{\tau}(x_{ij}), \ \mu_{\tau}=\mu_{\tau}(y_{i}), \ \sigma_{j\tau}=\sigma_{\tau}(x_{ij}) \mbox{ and } \sigma_{\tau}=\sigma_{\tau}(y_{i}).\) Under conditions \textbf{C1}-\textbf{C4}, we state the following result.

\begin{theorem}\label{thm:theo1}
When there are no influential observations in the dataset and \(\min(n,p) \rightarrow \infty,\) then, for any \(1 \leq k \leq n,\) and any sequence \((\tau_1,\cdots,\tau_q),\) we have \(\text{asym}T_{\min,k,m} \rightarrow_{d} \chi^{2}(1),\) and \(\text{asym}T_{\max,k,m} \rightarrow_{d} \chi^{2}(q)\) uniformly over \(m\) and \(\mathbb{A}_m=\lbrace A_r, 1\leq r \leq m \rbrace.\)
\end{theorem}

Notice that the number of asymmetric points \(q\) becomes the number of degrees of freedom of the chi-square distribution \(\chi^{2}(q).\)

\section{Simulation}\label{Sim}
We carried out simulations to assess the performance of the asymMIP measure with respect to bias, variable selection and sparsity. To this end, we generated datasets consisting of mostly non-influential observations (termed good observations) and as well as a small proportion of influential observations generated from a variety of  contamination schemes. 

\subsection{Design}\label{design}
We generated the good observations from equation model (\ref{eq:lm}). The predictor vector $(\boldsymbol{x}_i\transpose)$ was generated from a multivariate normal distribution with $\Cov(x_{ij}, x_{ij'})=0.5^{\lvert i-j \rvert}\ $ for $ \ 1\leq i \leq n, \ 1\leq j, j' \leq p$ with random error following a standard normal distribution. The parameter $\boldsymbol{\beta} = (0.3, \ 0.1, \ 0.2, \ 0.3, \ 0.9, \ 0.3, \ 1.1, \ 2.2, \ 0 , \ 0.4, \ 0, \ \ldots, \ 0)\transpose \in \mathbb{R}^p$ was assumed to be sparse. We set the sample size and the number of predictors to $n=250$ and $p= 1000,$ respectively. 

Then, the first $\tilde{n} = \floor{0.15 n} = 37$ of the good observations were replaced by influential observations,  generated from three different contamination schemes, following \citet{Zhao_2019}'s design. The influential observations are generated so that there will be either a masking effect (Model I)  a swamping effect (Model II), or both (Model III). 

In model I, masking was produced by generating influential observations that are clustered near each other. We selected the observation  with the maximum absolute value of the response variable as a seed observation. Then, we generated new predictor values for the influential observations near the seed observation, as follows:

\begin{equation*}
\widetilde{x}_{ij} = x_{i_0j} + \mathds{1}(j\in S_i)\times i/p, \qquad \widetilde{y}_i = y_{i_o} + \mu + \widetilde{\varepsilon}_i \times i/p, \qquad 1\leq j \leq p, \; 1\leq i \leq \widetilde{n},
\end{equation*}

where \(i_0=\argmax_{1\leq i \leq n} \lvert y_i \rvert\) and \(\widetilde{\varepsilon}_i\sim\mathcal{N}(0,\ 0.5).\) The values of the response variable for the influential observations, \(\ \widetilde{y}_i,\) depend on the parameter \(\mu \in \lbrace 4,\ 5,\ 6,\  7,\ 8,\ 9,\ 10 \rbrace\) which controls the degree of contamination, with larger values resulting in points that display more deviation. The set \(S_i, \ i=1,\ \ldots, \ \widetilde{n},\ \) is a subset of size \(10\) sampled independently with replacement from \(\lbrace 1, \ \ldots, \ p \rbrace.\)

In model II, we introduced influential observations with a swamping effect. We used a data generating process similar to equation model (\ref{eq:lm}), but designed to make influential observations look similar to good observations, as follows:

\begin{equation*}
\widetilde{y}_{i} = \sign(\sigma_i) \times (\widetilde{\boldsymbol{\beta}}\transpose\widetilde{\boldsymbol{x}}_{i} + \widetilde{\varepsilon}_i), \qquad 1\leq j \leq p, \; 1\leq i \leq \widetilde{n},
\end{equation*}

where \(\sigma_i\) is a binary variable with \(P(\sigma_i=1) = P(\sigma_i=-1)=1/2\) sampled independently of \((\widetilde{\boldsymbol{x}}_{i}, \ \widetilde{\varepsilon}_i).\) The parameter \(\widetilde{\boldsymbol{\beta}}\) is function of the initial parameter \(\boldsymbol{\beta}\) and is equal to: \(\widetilde{\boldsymbol{\beta}} = \boldsymbol{\beta} + (\mathbf{0}_{p-20}\transpose, \ \boldsymbol{w}\transpose),\) where \(\boldsymbol{w}=(w_1,\ \ldots, \ w_{20}) \in \mathbb{R}^{20}, \mbox{ and } w_j = j \times 0.005\mu.\) The predictors of the influential observations are generated by an independent normal distribution, \(\widetilde{\boldsymbol{x}}_{i} \sim\mathcal{N}\Big(\widetilde{\gamma},\ \mathbb{I}_p\Big),\) where \(\widetilde{\gamma} = (\mathbf{0}_{0.9p}\transpose, \ 0.5\mu \mathbf{1}_{0.1p}\transpose)\transpose.\) The strength of the association \(\widetilde{\boldsymbol{\beta}}\) depends on \(\mu,\) which ranges over \(\lbrace 4,\ 5,\ 6,\  7,\ 8,\ 9,\ 10\rbrace\) to control the severity of the swamping. Finally, we generated samples contaminated simultaneously by swamping and masking effects in Model III, where half of the influential observations are generated from model I and the other half from model II.

We estimated five variants of the asymMIP measure according to the expectiles: $\asymMIP_1$ with $\tau\in(0.25, \ 0.5, \ 0.75),\ \asymMIP_2$ with $\tau\in(0.3, \ 0.5, \ 0.7), \ \asymMIP_3$ with $\tau\in(0.4, \ 0.5, \ 0.6),\ \asymMIP_4$ with $\tau\in(0.25, \ 0.5, \ 0.6),$ and $\asymMIP_5$ with $\tau\in(0.4, \ 0.5, \ 0.75).$ We then compared the asymMIP measures with the MIP, asymHIM and HIM measures proposed by \citet{Zhao_2019}, \ \citet{zhao_high_dimensional_2013} and \citet{asymHimBarry2019}, respectively. All tests were performed using the Bonferroni procedure to control the family wise error at the nominal level \(\alpha=0.05.\) The simulation was carried out with 200 replications for each sample.

Note that we also generated contaminated samples according to the simulation design proposed by \citet{zhao_high_dimensional_2013} and \citet{asymHimBarry2019}. However, we only present the results related to the model where both the response and the predictor spaces are contaminated (refer to their paper for further details about the simulation design). The result of the other models is available upon request.

\subsection{Performance}\label{performance}

Performance was assessed through the (i) ability to identify influential observations with a small false positive rate, \((\FPR_{inf}).\) [Power], (ii) true positive rate \((\TPR_{inf}),\) i.e.\ the ability to identify the influential observations, and (iii) impact on subsequent analyses.

\begin{equation*}
  \TPR_{inf} = \frac{\lvert S_{inf}\cap\widehat{S}_{inf}\rvert}{n_{inf}} \quad \mbox{ and } \quad \FPR_{inf} = \frac{\lvert S_{inf}^{c}\cap\widehat{S}_{inf}\rvert}{n - n_{inf}},
\end{equation*}

where \(S_{inf}\) is the set of true influential observations, \(n_{inf}\) its size, and \(\widehat{S}_{inf}\) its estimator generated by an influence measure.

Then, we estimated the impact of including our influence measure in a data preprocessing pipeline prior to any statistical data analysis. We fitted the LASSO \citep{tibshiraniRegressionShrinkageSelection1996} on the initial raw data and then again on preprocessed data{\textemdash}data where influence measures have been applied to exclude potential influential observations. Then, we compared the results in term of bias, sparsity and predictive power. The bias was quantified as the distance between the true parameter and its estimator, and was represented by the following error function:

\begin{equation*}
  \ERR = \sqrt{\norm{\widehat{\boldsymbol{\beta}}-\boldsymbol{\beta}}_2}
\end{equation*}

where \(\boldsymbol{\beta}\) is the true parameter and \(\widehat{\boldsymbol{\beta}}\) its estimator, and \(\norm{\cdot}_2\) is the \(l_2\) norm.

The sparsity of the model was assessed by the true positive rate \((\TPR)\) and the false positive rate \((\FPR)\) for coefficient estimates. Let define the sets: \(\text{Supp}(\boldsymbol{\beta}) = \lbrace j \ | \ \beta_j \ne 0, \mbox{ for } j = 1,\dots,p \rbrace\) and \(\text{Supp}^c(\boldsymbol{\beta}) = \lbrace 1,\dots,p \rbrace \setminus \text{Supp}(\boldsymbol{\beta})\) then, the coefficient-based \(\TPR\) and \(\FPR\) are defined as:

\begin{equation*}
  \TPR = \frac{\lvert \text{Supp}(\boldsymbol{\beta})\cap\text{Supp}(\widehat{\boldsymbol{\beta}})\rvert}{\lvert \text{Supp}(\boldsymbol{\beta})\rvert}\quad \mbox{ and } \quad \FPR = \frac{\lvert \text{Supp}^c(\boldsymbol{\beta})\cap\text{Supp}(\widehat{\boldsymbol{\beta}})\rvert}{\lvert \text{Supp}^c(\boldsymbol{\beta})\rvert}.
\end{equation*}

Comparisons of results with  three other high dimensional influence measures MIP, asymHIM and HIM are shown with boxplots, with the addition of a straight line connecting the medians. We conducted a multiple comparison test to control the family wise error rate with respect to the subset sample size $(n_k, \ k=1,\ \ldots, \ m)$ using the Bonferroni test with a nominal level \(\alpha = 0.05\). The parameters of our algorithm were set to \(m=5\) and \(n_k=n/2\) for the number of subsets uniformly and randomly drawn with replacement and their size, respectively. The sensitivity and robustness of our influence measure according to this parameters is discussed below.

Simulations were conducted using high performance computing clusters provided by Calcul Quebec and Compute Canada, and using the R (v3.6.0) statistical programming language \citet{rcran}. We have implemented an R package, \textbf{hidetify}, is publicly available on GitHub at \url{https://github.com/AmBarry/hidetify}.

\subsection{Results}\label{results}

In this section, we present the performance of our asymMIP measure, while examining sensitivity and robustness with respect to the parameters in the algorithm. Finally, we discuss  tuning of the parameters of our algorithm.

\subsubsection{Performance of the influence measures in identifying the influential observations}

\begin{figure}
\centering
\includegraphics[width=\textwidth]{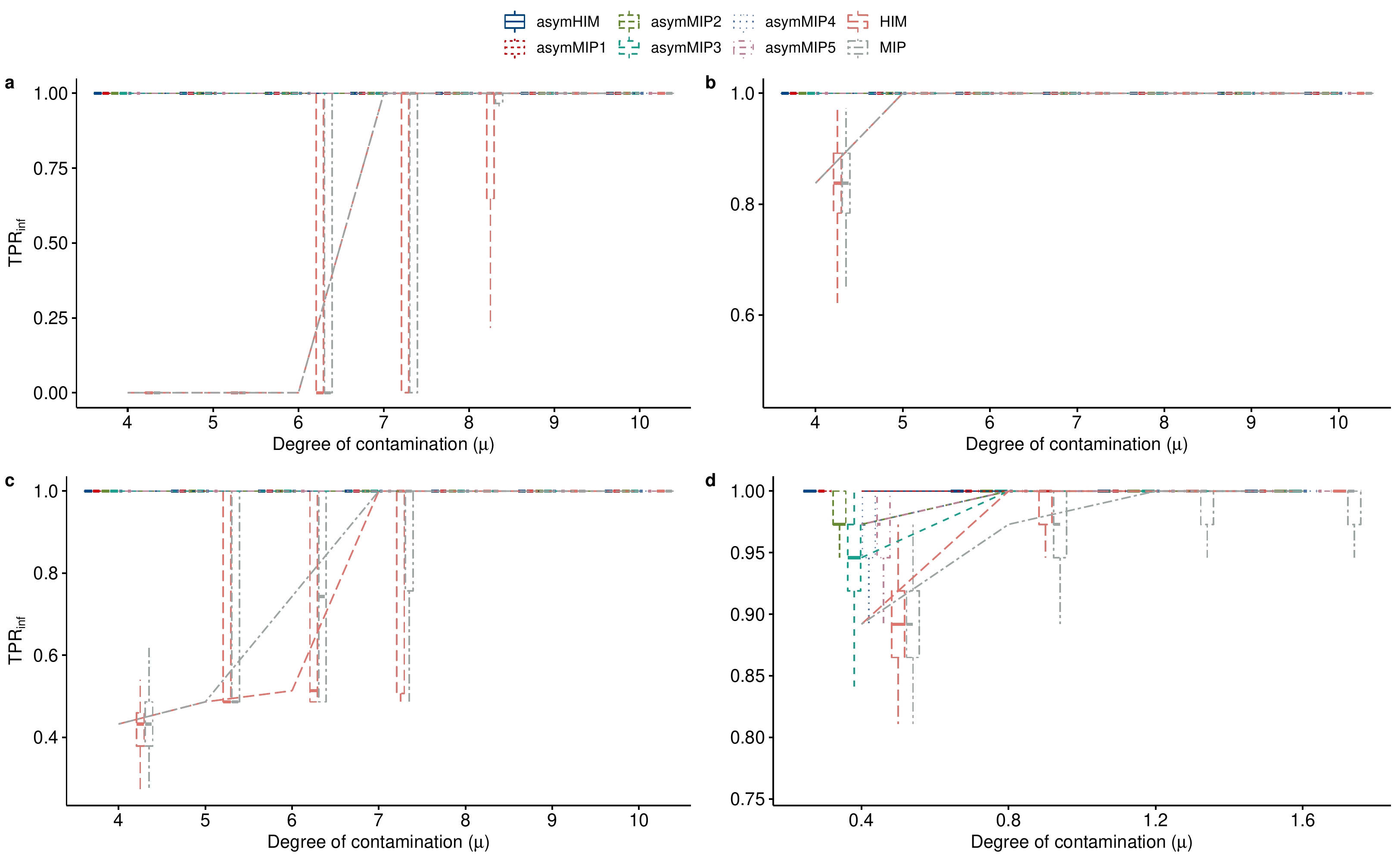}
\caption{Percentage of influential observations $(\TPR_{inf})$ identified by the different high dimensional influence measures $(\asymHIM, \ \asymMIP_1-\asymMIP_5, \ \HIM, \ \mbox{ and } \ \MIP).$ The data in Figure \ref{fig:power}\textbf{a}, Figure \ref{fig:power}\textbf{b} and Figure \ref{fig:power}\textbf{c} are generated by Model I, Model II and Model III; where $15\%$ of each sample is contaminated by masking, swamping and both, respectively. The data in Figure \ref{fig:power}\textbf{d} is generated by the third model proposed by \citet{zhao_high_dimensional_2013}.}\label{fig:power}%
\end{figure}

Figure \ref{fig:power} shows the proportion of influential observations \((\TPR_{inf})\) correctly identified as influential. The power \((\TPR_{inf})\) of the influence measures increases according to the degree of contamination $(\mu).$ All the new asymmetric influence measures proposed here $(\asymMIP_1 - \asymMIP_5)$ have similar performance (power close to 1) in all models, and outperform  the other influence measures. In contrast, the symmetric HIM and MIP measures have low power $(\TPR_{inf}< 50\%)$ for low-medium degrees of contamination $(\mu < 7),$ particularly in the presence of masking (Figure \ref{fig:power}\textbf{a}). 

\begin{figure}
\centering
\includegraphics[width=\textwidth]{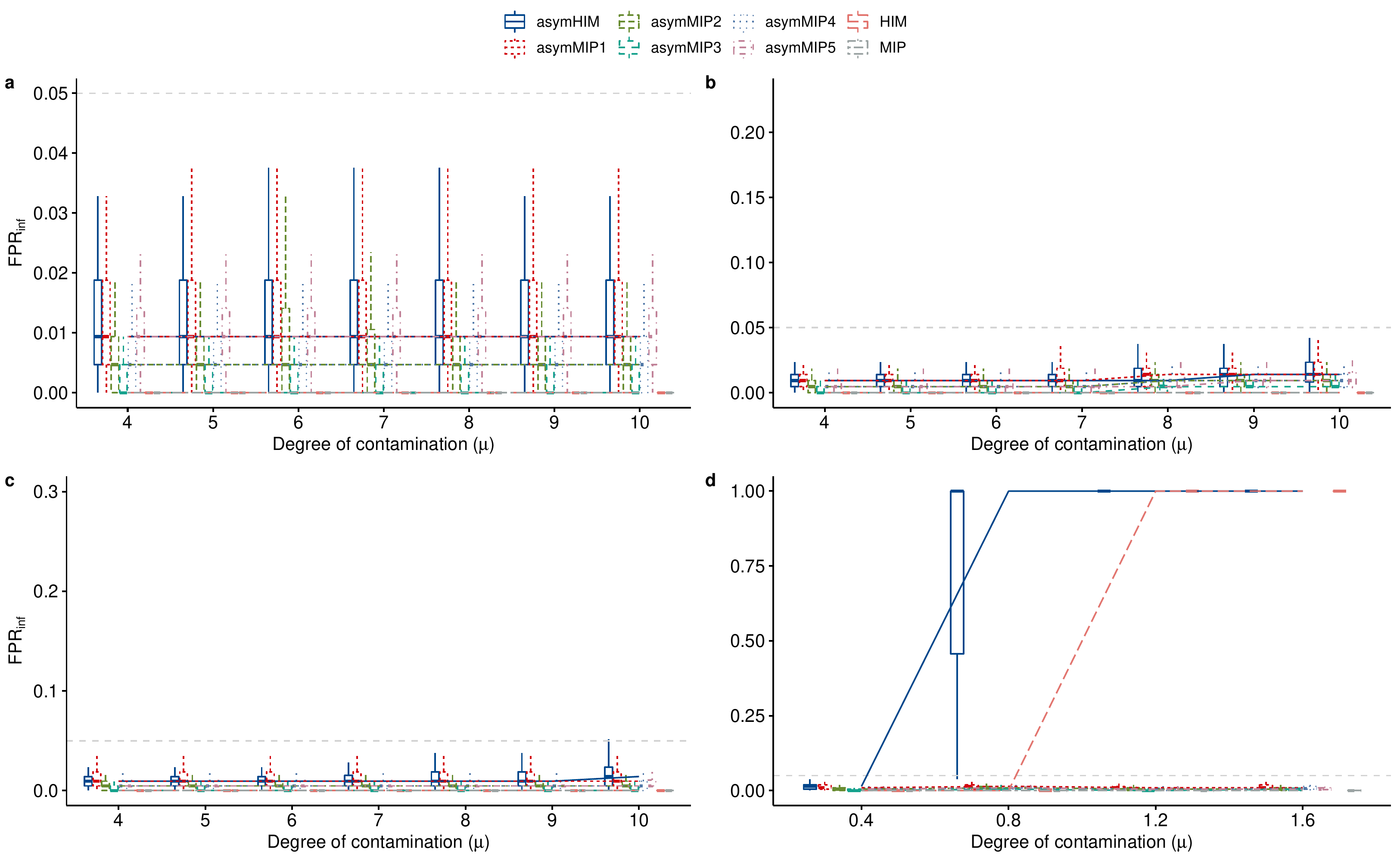}
\caption{Percentage of good observations falsely detected as influential $(\FPR_{inf})$ by the different high dimensional influence measures $(\asymHIM, \ \asymMIP_1-\asymMIP_5, \ \HIM, \ \mbox{ and } \ \MIP).$ The data in Figure \ref{fig:size}\textbf{a}, Figure \ref{fig:size}\textbf{b} and Figure \ref{fig:size}\textbf{c} are generated by Model I, Model II and Model III; where $15\%$ of each sample is contaminated by masking, swamping and both, respectively. The data in Figure \ref{fig:size}\textbf{d} is generated by the third model proposed by \citet{zhao_high_dimensional_2013}.}\label{fig:size}%
\end{figure}

Figure \ref{fig:size} displays the proportion of non-influential or good observations falsely identified as influential. The error rate \((\FPR_{inf})\) is below the \(5\%\) level for the algorithm-based influence measures $(\asymMIP_1 - \asymMIP_5 \mbox{ and MIP })$ in all models, except for the single detection techniques (asymHIM and HIM measures). Indeed, when the data is generated by the model proposed by \citet{zhao_high_dimensional_2013} and \citet{asymHimBarry2019}, we observe that HIM and asymHIM have a high error rate ($\FPR_{inf}$ close to \(100\%\)), when the degree of contamination $(\mu)$ is very high (Figure \ref{fig:size}\textbf{d}). We observe similar results when the number of predictors increases (Figure \sfref{fig:size_sample_pred} in the supplement). These two measures (HIM and asymHIM) are single detection techniques and are known to be highly affected by masking and swamping.

\subsubsection{Efficiency of the influence measures in subsequent analyses}

\begin{figure}
\centering
\includegraphics[width=\textwidth]{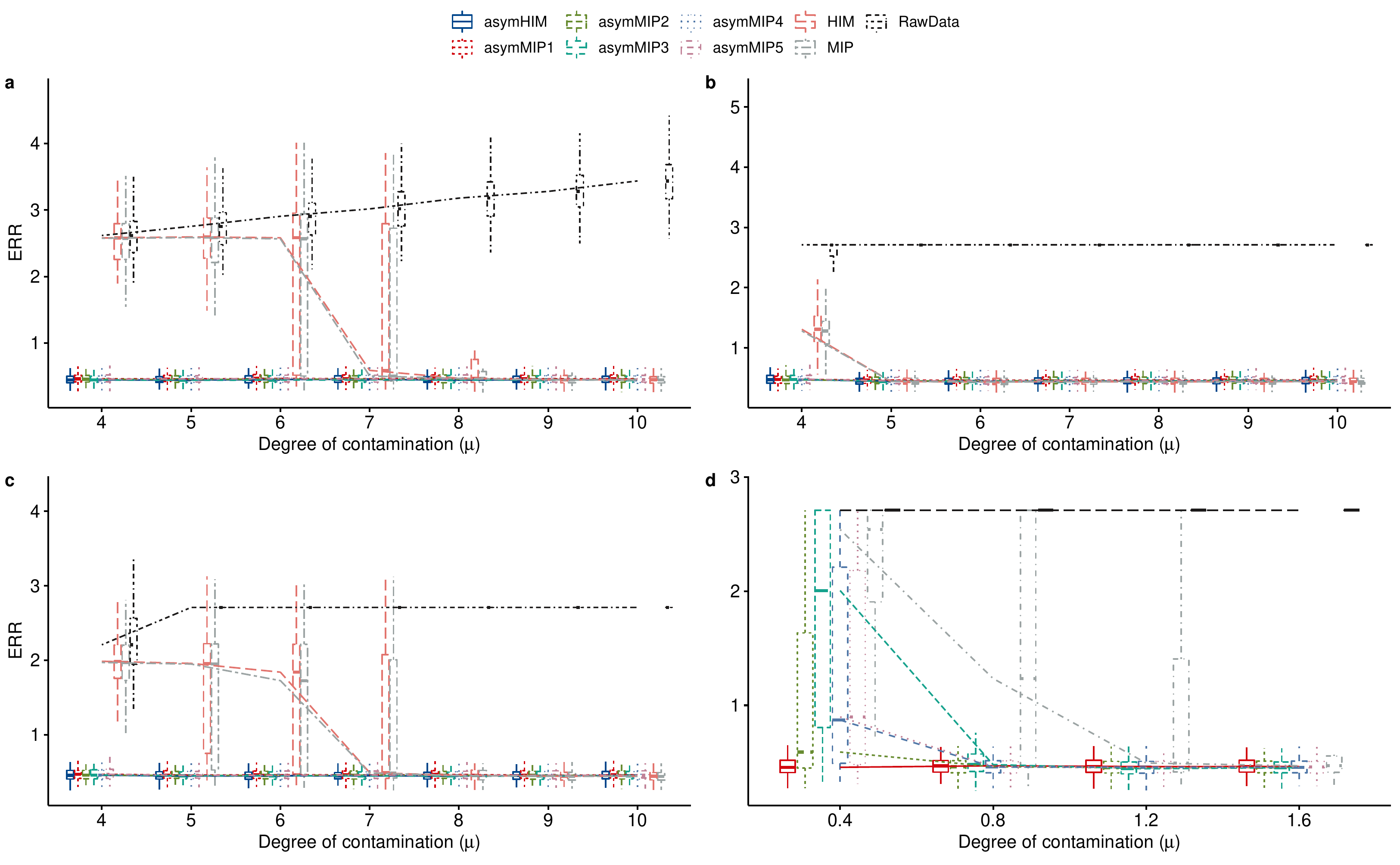}
\caption{The error or bias $(\ERR)$ that occurs in estimating the parameters of Model \\ref{eq:lm} from the contaminated 
data or data preprocessed by $(\asymHIM, \ \asymMIP_1-\asymMIP_5, \ \HIM, \ \mbox{ and } \ \MIP).$ 
The method $\rdata$ represents the bias resulting from using the initial contaminated raw data. 
The data in Figure \ref{fig:lasso_err}\textbf{a}, Figure \ref{fig:lasso_err}\textbf{b} and Figure \ref{fig:lasso_err}\textbf{c} 
are generated by Model I, Model II and Model III; where $15\%$ of each sample is contaminated by masking, swamping and both,
 respectively. The data in Figure \ref{fig:lasso_err}\textbf{d} is generated by the third model proposed by 
 \citet{zhao_high_dimensional_2013}.}\label{fig:lasso_err}%
\end{figure}

\begin{figure}
\centering
\includegraphics[width=\textwidth]{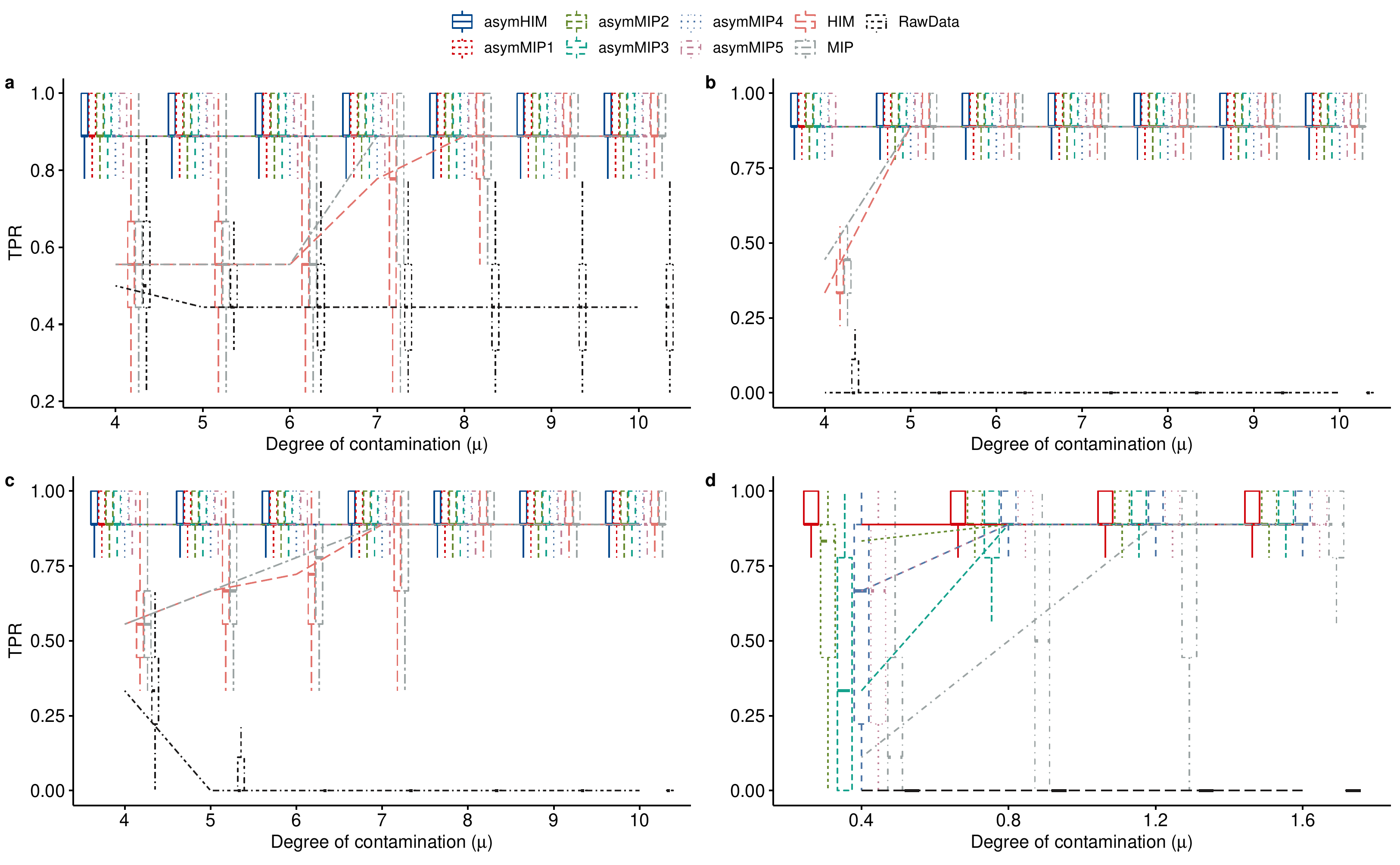}
\caption{Proportion of true non-zero effects detected $(\TPR)$, estimated with a non-zero effect size estimator from the contaminated data or data preprocessed by $(\asymHIM, \ \asymMIP_1-\asymMIP_5, \ \HIM, \ \mbox{ and } \ \MIP).$ The method $\rdata$ represents the $\TPR$ resulting from using the initial contaminated raw data. The data in Figure \ref{fig:lasso_tpr}\textbf{a}, Figure \ref{fig:lasso_tpr}\textbf{b} and Figure \ref{fig:lasso_tpr}\textbf{c} are generated by Model I, Model II and Model III; where $15\%$ of each sample is contaminated by masking, swamping and both, respectively. The data in Figure \ref{fig:lasso_tpr}\textbf{d} is generated by the third model proposed by \citet{zhao_high_dimensional_2013}.}\label{fig:lasso_tpr}%
\end{figure}

\begin{figure}
\centering
\includegraphics[width=\textwidth]{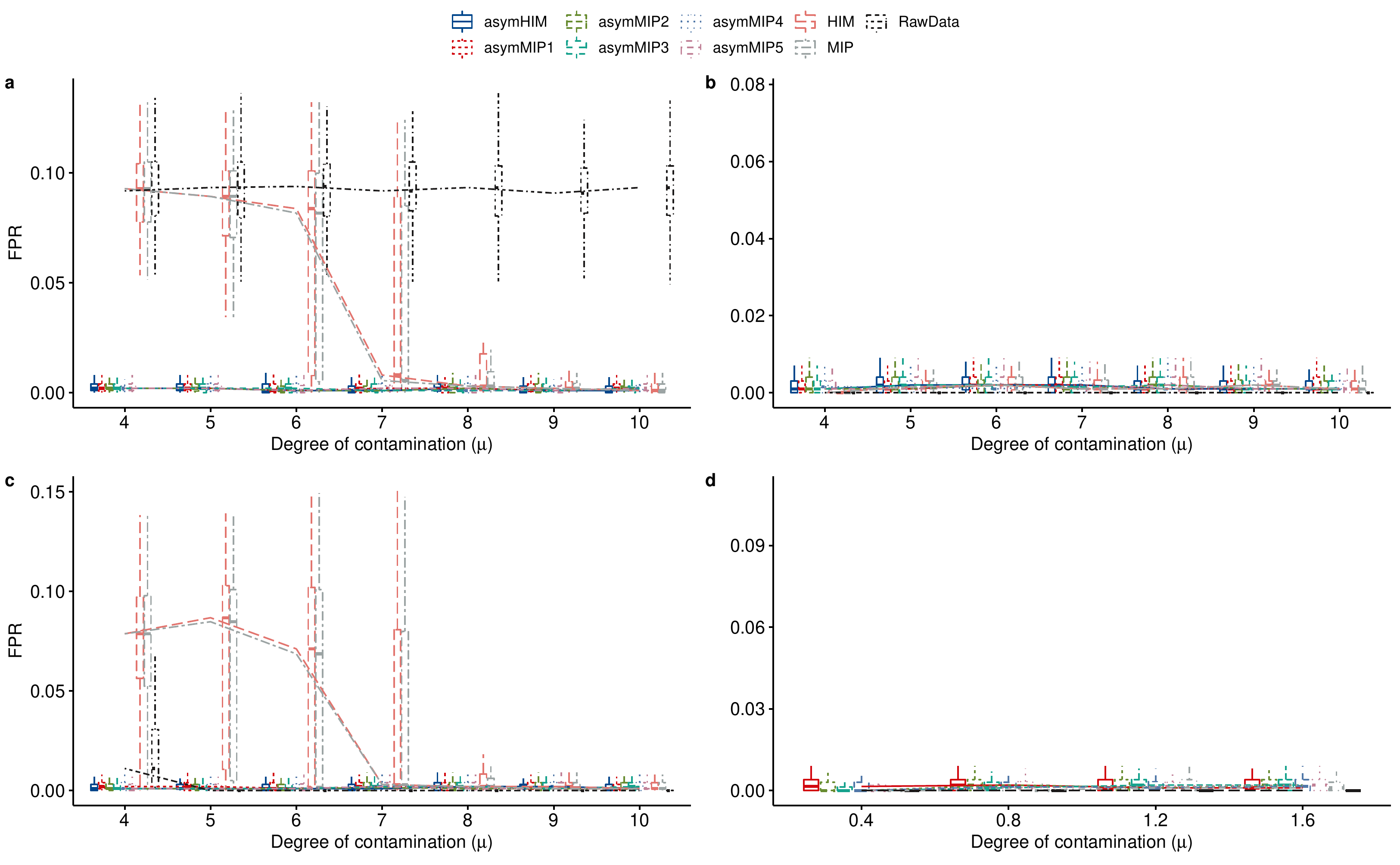}
\caption{Proportion of true zero coefficients $(\FPR)$ estimated with a non-zero effect size estimator from the contaminated data or data preprocessed by  $(\asymHIM, \ \asymMIP_1-\asymMIP_5, \ \HIM, \ \mbox{ and } \ \MIP).$ The method $\rdata$ represents the $\FPR$ resulting from using the initial contaminated raw data. The data in Figure \ref{fig:lasso_fpr}\textbf{a}, Figure \ref{fig:lasso_fpr}\textbf{b} and Figure \ref{fig:lasso_fpr}\textbf{c} are generated by Model I, Model II and Model III; where $15\%$ of each sample is contaminated by masking, swamping and both, respectively. The data in Figure \ref{fig:lasso_fpr}\textbf{d} is generated by the third model proposed by \citet{zhao_high_dimensional_2013}.}\label{fig:lasso_fpr}%
\end{figure}

Here, we report the effect of the influence measures on subsequent analyses. Figures \ref{fig:lasso_err}, \ref{fig:lasso_tpr} and \ref{fig:lasso_fpr} report the results of fitting the LASSO model to the data contaminated by influential observations and the data preprocessed by influence measures, with identified observations removed.
The $\rdata$ method in the figures corresponds to fitting the LASSO to the initial raw data including the influential observations. Figure \ref{fig:lasso_err} shows the error or bias \((\ERR)\) of the LASSO parameter estimator, where a small value of \(\ERR\) corresponds to a better estimate. Figure \ref{fig:lasso_tpr} shows the \(\TPR\) rate and Figure \ref{fig:lasso_fpr} reports the \(\FPR\) rate.

With the influential observations included, there is a deleterious impact on subsequent analyses. We observe that the $\rdata$ estimates display a large error (Figure \ref{fig:lasso_err}), captures a small fraction of the predictors with a non-zero effect size (Figure \ref{fig:lasso_tpr}), and falsely attributes a non-zero effect size estimate to several non-important predictors (Figure \ref{fig:lasso_fpr}). This leads to a biased, less sparse model with low predictive power. On the other hand, preprocessing the data by applying an influence measure, and particularly our influence measure (AsymMIP), before fitting the LASSO, leads to better results: lower bias, higher \(\TPR\) and lower \(\FPR.\) Our AsymMIP measure outperforms the MIP measure of  \citet{Zhao_2019} and  the single detection based influence measures (HIM and asymHIM). This highlights improvement associated with removing influential observations from the data. Note that the different asymMIP measures $(\asymMIP_1 - \asymMIP_5)$ with the expectiles selected within the first and third quartile have similar performance. 

\subsubsection{Sensitivity and robustness of the asymMIP influence measure}
We tested the sensitivity and robustness of the influence measures with respect to the sample size $(n\in\lbrace 100, \ 250, \ 500\rbrace),$ the number of predictor $(p\in\lbrace 1000, \ 3000, \ 5000\rbrace)$ and the proportion of influential observations \((\tilde{n} \in \lbrace 0\%n, \ 5\%n, \ 10\%n, \ 15\%n, \ 20\%n, \ \mbox{ and } \ 25\%n\rbrace).\) The results are available in the Supplement.

Our asymMIP influence measures outperform{\textemdash}have higher detection power \((\TPR_{inf})\) and an error size \((\FPR_{inf})\) below the \(5\%\) level, than the other influence measures regardless of the sample size and the number of predictors (Figure \sfref{fig:power_sample_pred} and Figure \sfref{fig:size_sample_pred} in the Supplement). For a fixed number of predictors (p), the power of the asymMIP influence measures increased, and their error decreased, with the sample size (n). In contrast, the power of the HIM and MIP influence measures decreased with respect to the sample size (n) (Figure \sfref{fig:power_sample_pred}). Note that the HIM and the asymHIM influence measures displayed an error rate higher than the $5\%$ level according to the degree of contamination $(\mu),$ particularly when the number of predictors is $p = 5000.$ Note that, these two measures (HIM and asymHIM) are single detection techniques and are known to be highly affected by masking and swamping.

Figure \sfref{fig:power_pct_outlier} and Figure \sfref{fig:size_pct_outlier} shows sensitivity and robustness of the influence measures with respect to the proportion of influential observations. Detection power \((\TPR_{inf})\) of asymMIP remains high even as the degree and proportion of contamination increase, except when the proportion of contamination in the data is \(\geq 25\% .\) In this case when the contamination proportion is very high, detection power decreases slightly (\sfref{fig:power_pct_outlier}); this also occurs when there is a low degree of contamination ($\mu=4$).
In contrast, the error rate \((\FPR_{inf})\) of the asymMIP measure decreases with the proportion of contamination in the data (Figure \sfref{fig:size_pct_outlier}). Also, the asymHIM and the $\asymMIP$ influence measures display an error rate \((\FPR_{inf})\) above the $5\%$ threshold when the data is not contaminated or is slightly contaminated (contamination below \(\leq 5\%).\) In contrast, the detection power of the HIM and MIP influence measures decreases substantially with the proportion of contamination in the data, Figure \sfref{fig:power_pct_outlier}.  

We also simulated data generated by an heteroscedastic and heavy tailed random error; results demonstrate the robustness of our asymMIP measure (Figure \sfref{fig:hetero} and Figure \sfref{fig:chi2}).

\subsubsection{Tuning the asymMIP algorithm parameters}

The algorithm that defines our asymMIP influence measure $(\ramm)$ depends on two main parameters: the number of subsets $(m)$ and their sample size $(n_k).$ We explored the sensitivity of the $\ramm$ algorithm (algorithm \ref{rmmmd_algo}) with respect to these parameters. The results showed that the power of detection of the asymMIP influence measure is still close to 1 and the false positive rate decreases with an increasing number of subsets $(m)$ (Figure \sfref{fig:number_subset}). We did not observe a clear pattern according to the sample size of the subsamples (Figure \sfref{fig:subset_size}).

\section{Real data application}\label{app}

Our methods are illustrated with data from the Autism Brain Imaging Data Exchange (ABIDE) neuroimaging dataset. The ABIDE initiative is an aggregation of functional and structural brain imaging data collected from laboratories around the world. Structural and functional neuroimaging datasets are high dimensional comprising voxel or vertex-wise measurements with strong spatial correlations, and are often contaminated due to image acquisition, and preprocessing artefacts \citep{fritsch_detecting_2012}. Indeed, the brain is a spatially embedded system, so there is a huge amount of spatial autocorrelation. This means that an influential observation will likely be extreme on many features. 

We applied the asymMIP $(\asymMIP_1-\asymMIP_5)$ multiple detection technique along with the MIP influence measure and the other single detection techniques (asymHIM and HIM) to the ABIDE neuroimaging dataset, and conducted a downstream analysis to predict brain maturity based on cortical thickness. Brain maturity prediction is a well studied subject in neuroscience to establish a baseline for normal brain development against which neurodevelopmental disorders can be assessed \citep{khundrakpam_prediction_2015}.

\textbf{Participants}

The ABIDE dataset comprises 573 control and 539 autism spectrum disorder (ASD) individuals from 16 international sites \citep{di_martino_autism_2014}. The neuroimaging data of these individuals were obtained from DataLad repository (\url{http://fcon_1000.projects.nitrc.org/indi/abide/}). Only a subset of individuals is used in this work due to failures resulting from the image processing pipeline. A demographic description of these individuals is provided by \citet{bhagwat2021}. In this paper, we only used the 542 control individuals from the ABIDE dataset.

\textbf{MR Image processing and cortical thickness measurements}

The MR images (T1-weighted scans) were processed using the FreeSurfer 6.0 pipeline (Fischl 2012) deployed on CBrain - a high-performance computing facility \citep{sherif_cbrain_2014}. FreeSurfer delineates the cortical surface from a given MR scan and quantifies thickness measurements on this surface for each brain hemisphere. The cortical thickness measurements for each MR image are computed using FreeSurfer 6.0 pipeline \citep{fischl_freesurfer_2012, dale_cortical_1999}. The pipeline consists of 1) affine registration to the MNI305 space \citep{collins_automatic_1994}; 2) bias field correction; 3) removal of skull, cerebellum, and brainstem regions from the MR image; 4) estimation of white matter surface based on MR image intensity gradients between the white and gray matter; and 4) estimation of pial surface based on intensity gradients between the gray matter and cerebrospinal fluid (CSF). The distance between the white and pial surfaces provides the thickness estimate at a given location of cortex. For detailed descriptions refer to \citet{fischl_freesurfer_2012} and \citet{dale_cortical_1999}. The individual cortical surfaces are then projected onto a common space (i.e.~fsaverage) characterized by 163,842 vertices per hemisphere to establish inter-individual spatial correspondence.

\textbf{Data cleaning and Age prediction}

\begin{figure}
     \centering
     \begin{subfigure}[b]{0.45\textwidth}
         \centering
         \includegraphics[width=\textwidth]{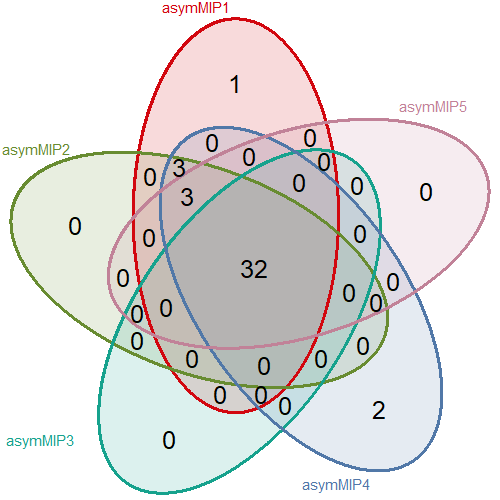}
         \caption{}
         \label{fig:venn_diagramm_asym_mip}
     \end{subfigure}
     \hfill
     \begin{subfigure}[b]{0.45\textwidth}
         \centering
         \includegraphics[width=\textwidth]{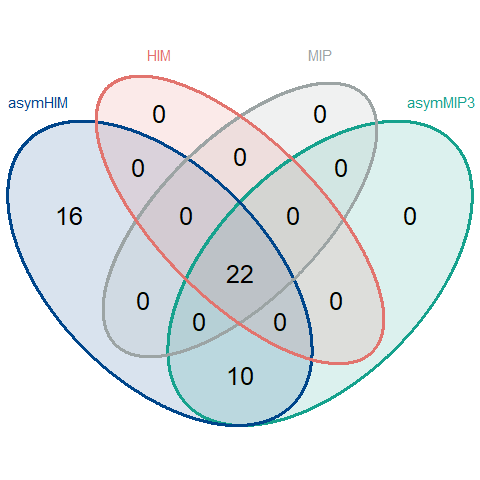}
         \caption{}
         \label{fig:venn_diagramm_outlier_set}
     \end{subfigure}
       \caption{Number of influential observations identified in the ABIDE dataset. Different colours are used for each influence measure. Venn diagram \textbf{(a)} displays the number of influential observations identified by the asymmetric influence measures $(\asymMIP_1-\asymMIP_5)$ which identified 32 influential observations in common. Venn diagram \textbf{(b)} displays the number of influential observations identified by the influence measures $(\asymHIM, \ \asymMIP_3, \ \HIM, \ \mbox{ and } \ \MIP)$ which identified 22 influential observations in common.}
        \label{fig:two_venn}
\end{figure}

We used the chronological age (i.e. brain maturity) of the patients in the control group as a response variable in the multiple detection procedure. The influential observations identified by the different influence measures are presented in Figure \ref{fig:two_venn}. The asymmetric asymMIP $(\asymMIP_1-\asymMIP_5)$ influence measures jointly identified 32 influential observations, the $\asymMIP_4$ influence measure identified 40 influential observations and the $\asymMIP_3$ influence measure identified 32. We have conservatively chosen the smallest set of influential observations (32) as the set of influential observations to reduce the false positive impact. The asymHIM influence measure captures the largest influential set (48) and in contrast, the HIM and MIP influence measures reported the smallest set of influential observations (22) (Figure \ref{fig:two_venn}). These results are in agreement with the defining characteristics of these measures:  asymHIM is aggressive and tends to identify all the influential observations with a high false positive rate, wherease  HIM and  MIP are conservative with low detection power and a very low false positive rate. Our new algorithm-based multiple detection influence measure (asymMIP) is a reasonable compromise between the aggressive  (asymHIM) and conservative (HIM and MIP) influence measures .

\begin{figure}
\centering
\includegraphics[width=\textwidth]{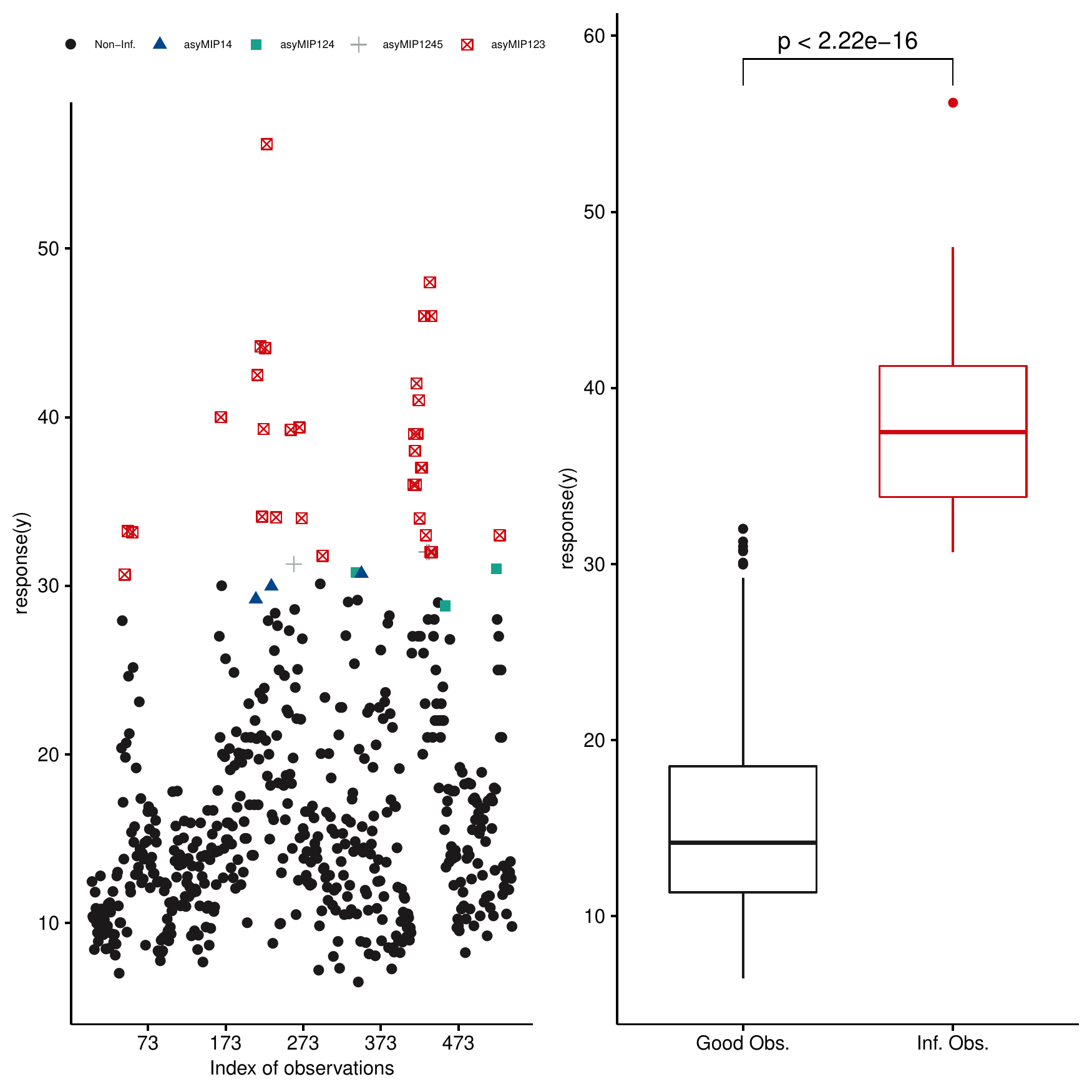}
\caption{The left figure displays a scatterplot of chronological age and influential status, as labelled by $(\asymMIP_1-\asymMIP_5).$ The figure in the right represents the chronological age distribution of the non-influential observations (Good Obs.) and the 32 influential observations (Inf. Obs.) identified by the asymmetric influence measures. The Kruskal–Wallis test compares the two samples (p-value).}\label{fig:scatter_boxplot_age}%
\end{figure}

\begin{figure}
\centering
\includegraphics[width=\textwidth]{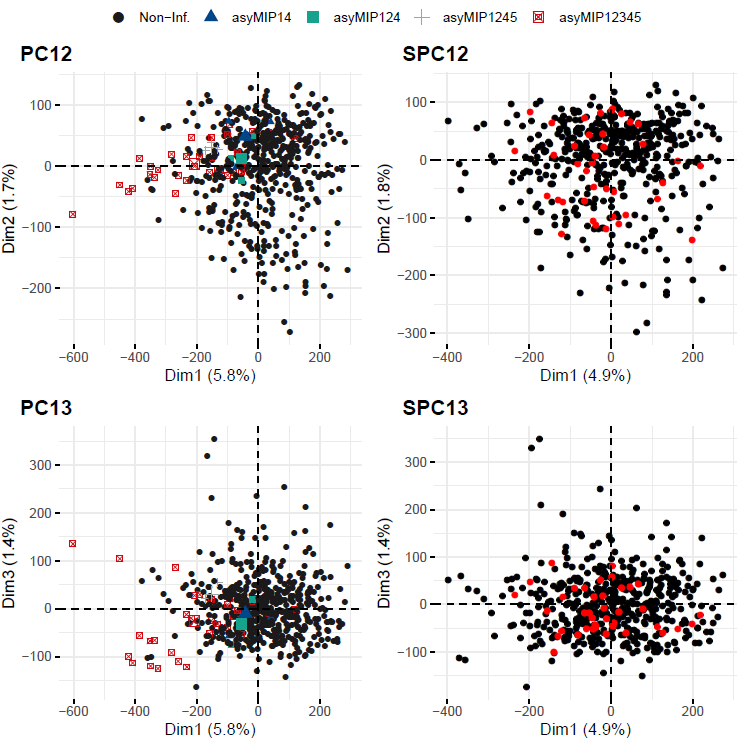}
\caption{Score plots of the observations in the first three dimensions of the principal component analysis (PCA) according to their influential nature identified by  $(\asymMIP_1-\asymMIP_5).$ 
The influential observations contributed to the calculation of the PCs in figures \textbf{PC12} and \textbf{PC13}. However, they did not contribute to the variance of the PCs in figures \textbf{SPC12} and \textbf{SPC13}; where these points are projected and displayed in red.}\label{fig:pca_supp_pca12_13}%
\end{figure}

In order to visualize age with the cortical thickness and the influential points, we performed a principal component analysis (PCA) of the cortical thickness measurements and display the first PC in Figure \ref{fig:scatter_boxplot_age} and Figure \ref{fig:pca_supp_pca12_13}. These scatterplots show that the influential observations all came from subjects with age over 30, and the age distributions of the influential and good observations are statistically different, Figure \ref{fig:scatter_boxplot_age}. The first three PCs explain only a small percentage of the variance $(9\%),$ suggesting that more PCs are probably  necessary for a better visualization and representation of the data variation. Nevertheless, the first PC does discriminate between the core of the influential observations and assumed good observations (Figure \ref{fig:pca_supp_pca12_13}).

After the data cleaning process, we applied a supervised machine learning model to predict the chronological age based on cortical thickness measurements. In this regard, we followed \citet{khundrakpam_prediction_2015}'s approach. We estimated a Elastic Net model \citep{Friedman2010} with a hyper-parameter $\theta=0.5,$ balancing the \(L_1\) and \(L_2\) penalization. We applied a 10-fold nested cross-validation (CV) loops to simultaneously select the best regularization parameter and evaluate the prediction accuracy \citep{ambroise_selection_2002}. We selected the best regularization parameter in the inner 10-fold CV loop and we evaluated the age prediction accuracy in the outer 10-fold CV loop. We used the mean absolute error (MAE) between the chronological and the estimated age in the inner loop, and the correlation coefficient between the chronological and estimated age in the outer loop as a measure of goodness of fit. Finally, we repeated the analysis 100 times and reported the results in the form of distribution using violin plots. The age response is predicted based on 299,569 cortical vertex features, 149,778 cortical vertices from the right hemisphere, and 149,791 cortical vertices from the left hemisphere. The method $\rdata$ represent the results of the model fitted to the initial data including all the samples.

\textbf{Results}

\begin{figure}
\centering
\includegraphics[width=\textwidth]{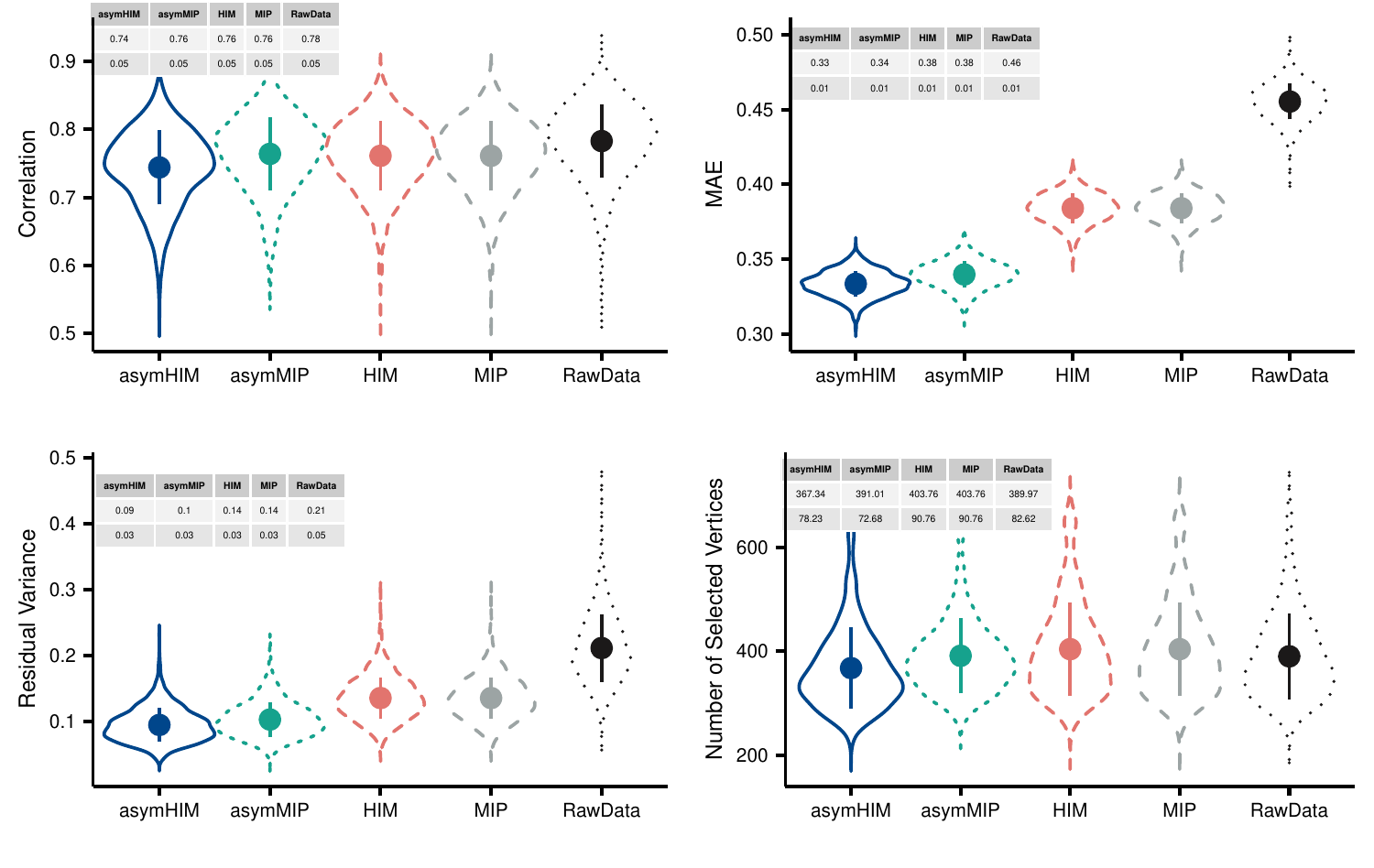}
\caption{Elasticnet results for the ABIDE dataset. Violin plots show correlation between chronological and predicted age (top left), mean absolute error (MAE, top right), residual variance between chronological and predicted age (bottom left), and number of selected cortical vertices (bottom right) across 100 repeated nested CV and detection methods (asymHIM, \ asymMIP, \ HIM, \ MIP, \ RawData).}\label{fig:glmnet_main_result}%
\end{figure}

The results on the ABIDE data in  Figure \ref{fig:glmnet_main_result} show that the selection model performs better (low MAE and low variance) on the cleaned data than on the uncleaned data. These performance measures are even better with the data cleaned by our asymMIP influence measure. Indeed, we observe that the selection model displays a mean MAE of 0.46 and a mean variance of 0.21 with the uncleaned data. While these values are 0.34 and 0.1, respectively when the selection model is fitted to the data cleaned by the asymMIP influence measure. 

The number of vertices selected by the model is on average slightly higher with the cleaned data than with the uncleaned data. This is not surprising, because the Elastic Net model $(\theta=0.5)$ inherits the properties of the Ridge regression model and will have a tendency to shrink the correlated vertices toward each other. Correlation between the chronological age and the predicted age is slightly higher with the uncleaned data (mean correlation = 0.78) than the cleaned data (mean correlation = 0.76); this lower correlations is likely a consequence of the fact that the influential observations were associated with older individuals and therefore the variance in age in the dataset is decreased by removal of the identified influential points. 
Indeed, these measures are correlation based, and observations with high marginal correlations have higher probability to be identified as influential and removed from the dataset. 

Note that the asymMIP results presented here (Figure \ref{fig:glmnet_main_result}) correspond to the $\asymMIP3$ influence measure that identified the smallest set of influential observations (33 in total). However, the selection models applied to the data cleaned by the different asymMIP $(\asymMIP_1-\asymMIP_5)$ influence measures yielded similar results (Figure \sfref{fig:glmnet_all_asym} in the supplement). 

\begin{figure}
\centering
\includegraphics[width=\textwidth]{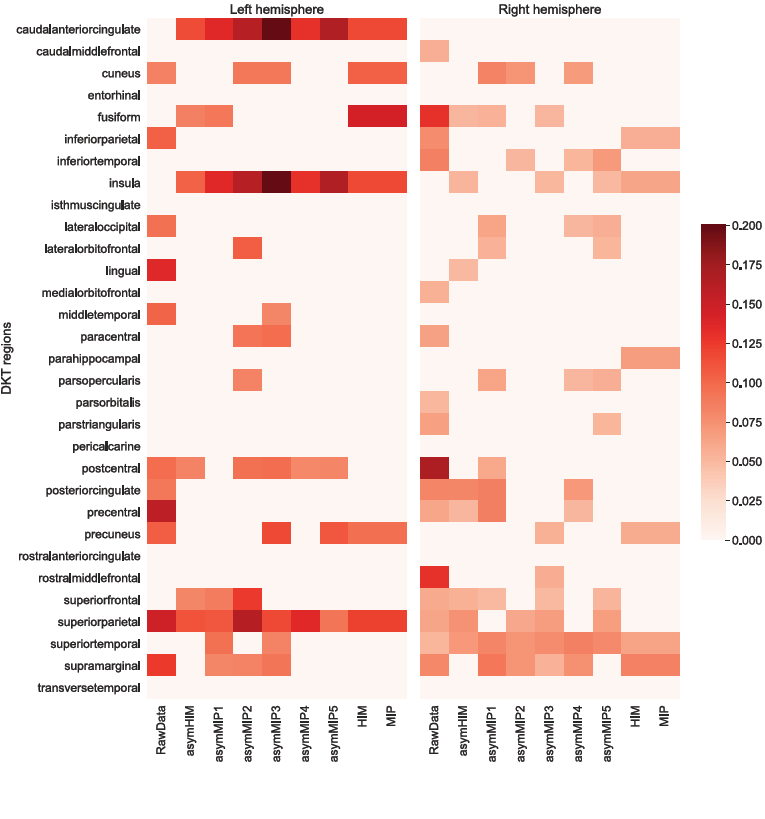}
\caption{Heatmap displaying top regions of interest (ROIs) for predicting chronological age. The ROIs are defined by the Desikan-Killiany-Tourville cortical parcellation comprising 31 bilateral regions mapped on the FreeSurfer surface mesh. The important ROIs are selected according to the absolute effect size of their parameter estimate for the member vertices.}\label{fig:heatmap_full}%
\end{figure}

We then examined regions identified as important for brain maturity  according to the absolute effect size of their parameter estimate (Figure \ref{fig:heatmap_full}). We used the Desikan-Killiany-Tourville (DKT) cortical parcellation with 31 bilateral regions for anatomical mapping of Free Surfer surface mesh \citep{klein101LabeledBrain2012}. 

According to the heatmap in Figure \ref{fig:heatmap_full}, all selection models, including the selection model fitted to the uncleaned data $(\rdata),$ identified regions such as the superior parietal (left hemisphere), or superior temporal (right hemisphere) as regions with vertices that are top predictors of the chronological age. However, we observed that the insula and the caudal anterior cingulate regions in the left hemisphere are particularly highlighted by the cleaned data based selection models. Almost all these regions, and most of the regions identified in the heatmap, are also reported as regions predictive of the chronological age by \citet{khundrakpam_prediction_2015}. 

\section{Discussion}\label{discus}

We have proposed an algorithm-based influence measure to identify influential observations in high dimensional regression that copes with masking and swamping.  Our method builds on expectiles to capture influential points even in the presence of heterogeneous data distributions, and therefore has a higher detection power than other methods. Availability of asymptotic distributions for the statistics involved in the $\ramm$ algorithm avoids the computational challenge that a bootstrap would have induced, and facilitates systematic application of our influence measure during quality control and preprocessing.

First, our simulations showed that the asymMIP influence measure is more effective at identifying influential observations in all scenarios than the other three methods (asymHIM, HIM, MIP). Second, simulations also showed that inclusion of our asymMIP influence measure in preprocessing led to less biased estimators and a better predictive model. In contrast, failing to remove the influential observations, partially or completely led to less than ideal results: estimates with more bias and models with lower predictive power. Moreover, the asymMIP influence measure is robust to changes in sample size, number of predictors,  proportion of contamination, and to the presence of heteroscedasticity or heavy tailed random error. 

After preprocessing  the ABIDE data with these influence measures, we were able to reveal interesting insights into the prediction of brain maturity based on cortical thickness. We identified the $30+$ age group as having different influence than the younger age group on the prediction of brain maturity. Furthermore,  application of the influence measures improved the accuracy of the selection models while preserving their predictive power. Using asymMIP for preprocessing led to the model with the best accuracy.

Attention must be drawn to the parameters of the $\ramm$ algorithm: the number of subsets drawn uniformly and randomly with replacement and their size in the Min and Max steps. These two parameters are related to finding an uncontaminated sample. Simulations showed that sampling a larger number of subsets contributed to a reduction in the error rate of asymMIP while maintaining the detection power. In contrast, sensitivity analysis did not reveal dependence on size of these subsets. Since both parameters depend on the sample size, we suggest setting the same default values that we applied in our simulation{\textemdash}5 subsets of size $n/2,$ when the sample size is small. When the sample size is large then there is more flexibility to increase the number of subsets and to vary their size. 

Our $\ramm$ algorithm also depends on the sequence of $\tau$ values for the expectiles. In  \citet{asymHimBarry2019}, we showed that the false positive rate increased with the length of the expectile sequence, and that a sequence of length three seemed to be a good compromise between detection power and false positives. For example, an influence measure based on a sequence of two expectiles displayed a lower false positive rate than asymMIP (which is based on three expectiles), and a higher detection rate than MIP (which used only one, the median). However, the influence measure{\textemdash}influence measure using two expectiles will not capture $100\%$ of the influential observations in all scenarios. There is always a trade off between the detection power and the false positive rate. Nevertheless, based on the second part of our simulation study, we have shown that it is more advantageous to have an influence measure with high power and a slightly high false positive rate than an influence measure with a lower false positive rate and also low power, since failing to remove all influential observation has more adverse effects. 

It is important to note the slightly high error rate of the asymMIP influence measure when the data is not contaminated at all. This behaviour arises from  aggressive statistic in the  \textit{MAX step} of the $\ramm$ algorithm, which is effective at addressing masking, but overly sensitive when the data is not contaminated. To alleviate this problem, we suggest testing the null hypothesis in the \textit{MAX step} of the $\ramm$ algorithm at a lower threshold, e.g. $0.1\%.$ Hence, the $\ramm$ algorithm contains three different parameters to test each null hypothesis in the three steps  of the algorithm. In addition to asymMIP with three expectiles, we recommend also estimate $\ramm$ with one expectile ($\tau=0.5$) corresponding to MIP, since this is a good proxy for assessing the degree of contamination of the data.

Multivariate or unsupervised high-dimensional models are not addressed by asymMIP, and would also benefit from development and implementation of appropriate diagnostic measures. Despite the  computational challenges that are involved, development of novel influence measures for high-dimensional data is an active and growing field of research of great importance. 

\section{Conclusion}\label{con}
In this paper, we proposed an algorithm-based influence measure  that is not misled by swamping and masking effects in high dimensional regression. Our three-step algorithm employs a conservative statistic in the first step to minimize swamping effects, then an aggressive statistic in the second step to overcome masking. In the last step, the single detection technique is applied to validate the set of influential observations. The derivation of the asymptotic distributions of the different statistics involved in the algorithm avoids the computational burden that a bootstrap would have induced. Thus, our algorithm, implemented in an R package called hidetify, and publicly available at \url{github.com/AmBarry/hidetify}, is computationally efficient, an important consideration for high-dimensional data. The integration of our influence measure in  quality control and preprocessing pipelines will minimize any deleterious impact of influential observations in downstream analysis.

\clearpage
\bibliographystyle{apalike}
\bibliography{ref_mip.bib}

\begin{thebibliography}{}

\bibitem[Ambroise and McLachlan, 2002]{ambroise_selection_2002}
Ambroise, C. and McLachlan, G.~J. (2002).
\newblock Selection bias in gene extraction on the basis of microarray
  gene-expression data.
\newblock {\em Proceedings of the National Academy of Sciences},
  99(10):6562--6566.

\bibitem[Barry et~al., 2020a]{asymHimBarry2019}
Barry, A., Bhagwat, N., Misic, B., Poline, J.-B., and Greenwood, C. M.~T.
  (2020a).
\newblock Asymmetric influence measure for high dimensional regression.
\newblock {\em Communications in Statistics - Theory and Methods}, 0(0).

\bibitem[Barry et~al., 2020b]{GEEE_Barry2018}
Barry, A., Oualkacha, K., and Charpentier, A. (2020b).
\newblock A new {GEE} method to account for heteroscedasticity, using
  asymmetric least-square regressions.
\newblock {\em arXiv:1810.09214 [stat]}.
\newblock arXiv: 1810.09214.

\bibitem[Belsley et~al., 1980]{belsley_regression_1980}
Belsley, D.~A., Kuh, E., and Welsch, R.~E. (1980).
\newblock {\em Regression diagnostics: identifying influential data and sources
  of collinearity}.
\newblock Wiley.

\bibitem[Bhagwat et~al., 2021]{bhagwat2021}
Bhagwat, N., Barry, A., Dickie, E.~W., Brown, S.~T., Devenyi, G.~A., Hatano,
  K., DuPre, E., Dagher, A., Chakravarty, M., Greenwood, C. M.~T., Misic, B.,
  Kennedy, D.~N., and Poline, J.-B. (2021).
\newblock Understanding the impact of preprocessing pipelines on neuroimaging
  cortical surface analyses.
\newblock {\em GigaScience}, 10(giaa155).

\bibitem[Chatterjee and Hadi, 1986]{chatterjee_influential_1986}
Chatterjee, S. and Hadi, A.~S. (1986).
\newblock Influential {Observations}, {High} {Leverage} {Points}, and
  {Outliers} in {Linear} {Regression}.
\newblock {\em Statistical Science}, 1(3):379--393.

\bibitem[Collins et~al., 1994]{collins_automatic_1994}
Collins, D.~L., Neelin, P., Peters, T.~M., and Evans, A.~C. (1994).
\newblock Automatic {3D} intersubject registration of {MR} volumetric data in
  standardized talairach space.
\newblock {\em Journal of Computer Assisted Tomography}, 18(2):192--205.

\bibitem[Dale et~al., 1999]{dale_cortical_1999}
Dale, A., Fischl, B., and Sereno, M. (1999).
\newblock Cortical surface-based analysis. i. segmentation and surface
  reconstruction.
\newblock {\em NeuroImage}, 9(2):179—194.

\bibitem[Di~Martino et~al., 2014]{di_martino_autism_2014}
Di~Martino, A., Yan, C.-G., Li, Q., Denio, E., Castellanos, F.~X., Alaerts, K.,
  Anderson, J.~S., Assaf, M., Bookheimer, S.~Y., Dapretto, M., Deen, B.,
  Delmonte, S., Dinstein, I., Ertl-Wagner, B., Fair, D.~A., Gallagher, L.,
  Kennedy, D.~P., Keown, C.~L., Keysers, C., Lainhart, J.~E., Lord, C., Luna,
  B., Menon, V., Minshew, N.~J., Monk, C.~S., Mueller, S., Müller, R.-A.,
  Nebel, M.~B., Nigg, J.~T., O'Hearn, K., Pelphrey, K.~A., Peltier, S.~J.,
  Rudie, J.~D., Sunaert, S., Thioux, M., Tyszka, J.~M., Uddin, L.~Q.,
  Verhoeven, J.~S., Wenderoth, N., Wiggins, J.~L., Mostofsky, S.~H., and
  Milham, M.~P. (2014).
\newblock The autism brain imaging data exchange: towards a large-scale
  evaluation of the intrinsic brain architecture in autism.
\newblock {\em Molecular psychiatry}, 19(6):659--67.

\bibitem[Diakonikolas et~al., 2021]{Diakonikolas2021}
Diakonikolas, I., Kamath, G., Kane, D.~M., Li, J., Moitra, A., and Stewart, A.
  (2021).
\newblock Robustness meets algorithms.
\newblock {\em Commun. ACM}, 64(5):107–115.

\bibitem[Fan and Lv, 2008]{fan_sure_2008}
Fan, J. and Lv, J. (2008).
\newblock Sure independence screening for ultrahigh dimensional feature space.
\newblock {\em Journal of the Royal Statistical Society: Series B (Statistical
  Methodology)}, 70(5):849--911.

\bibitem[Filzmoser et~al., 2008]{filzmoser_outlier_2008}
Filzmoser, P., Maronna, R., and Werner, M. (2008).
\newblock Outlier identification in high dimensions.
\newblock {\em Computational Statistics \& Data Analysis}, 52(3):1694--1711.

\bibitem[Fischl, 2012]{fischl_freesurfer_2012}
Fischl, B. (2012).
\newblock {FreeSurfer}.
\newblock {\em NeuroImage}, 62(2):774 -- 781.

\bibitem[Friedman et~al., 2010]{Friedman2010}
Friedman, J.~H., Hastie, T., and Tibshirani, R. (2010).
\newblock Regularization paths for generalized linear models via coordinate
  descent.
\newblock {\em Journal of Statistical Software, Articles}, 33(1):1--22.

\bibitem[Fritsch et~al., 2012]{fritsch_detecting_2012}
Fritsch, V., Varoquaux, G., Thyreau, B., Poline, J.-B., and Thirion, B. (2012).
\newblock Detecting outliers in high-dimensional neuroimaging datasets with
  robust covariance estimators.
\newblock {\em Medical Image Analysis}, 16(7):1359--1370.

\bibitem[Jeong and Kim, 2018]{jeong_effect_2018}
Jeong, J. and Kim, C. (2018).
\newblock Effect of outliers on the variable selection by the regularized
  regression.
\newblock {\em Communications for Statistical Applications and Methods},
  25(2):235--243.

\bibitem[Karoui et~al., 2013]{NoureddineElKaroui2013}
Karoui, N.~E., Bean, D., Bickel, P.~J., Lim, C., and Yu, B. (2013).
\newblock On robust regression with high-dimensional predictors.
\newblock {\em Proceedings of the National Academy of Sciences of the United
  States of America}, 110(36):14557--14562.

\bibitem[Khundrakpam et~al., 2015]{khundrakpam_prediction_2015}
Khundrakpam, B.~S., Tohka, J., Evans, A.~C., and {Brain Development Cooperative
  Group} (2015).
\newblock Prediction of brain maturity based on cortical thickness at different
  spatial resolutions.
\newblock {\em NeuroImage}, 111:350—359.

\bibitem[Klein and Tourville, 2012]{klein101LabeledBrain2012}
Klein, A. and Tourville, J. (2012).
\newblock 101 {Labeled} {Brain} {Images} and a {Consistent} {Human} {Cortical}
  {Labeling} {Protocol}.
\newblock {\em Frontiers in Neuroscience}, 6.

\bibitem[Loh, 2017]{PoLingLoh2017}
Loh, P.-L. (2017).
\newblock Statistical consistency and asymptotic normality for high-dimensional
  robust m-estimators.
\newblock {\em The Annals of Statistics}, 45(2):866--896.

\bibitem[Maronna, 2011]{Maronna2011}
Maronna, R.~A. (2011).
\newblock Robust ridge regression for high-dimensional data.
\newblock {\em Technometrics}, 53(1):44--53.

\bibitem[Newey and Powell, 1987]{newey_asymmetric_1987}
Newey, W.~K. and Powell, J.~L. (1987).
\newblock Asymmetric least squares estimation and testing.
\newblock {\em Econometrica}, 55(4):819--47.

\bibitem[Nurunnabi et~al., 2014]{nurunnabi_procedures_2014}
Nurunnabi, A., Hadi, A.~S., and Imon, A. (2014).
\newblock Procedures for the identification of multiple influential
  observations in linear regression.
\newblock {\em Journal of Applied Statistics}, 41(6):1315--1331.

\bibitem[Prasad et~al., 2020]{Prasad2020}
Prasad, A., Suggala, A.~S., Balakrishnan, S., and Ravikumar, P. (2020).
\newblock Robust estimation via robust gradient estimation.
\newblock {\em Journal of the Royal Statistical Society Series B},
  82(3):601--627.

\bibitem[{R Core Team}, 2021]{rcran}
{R Core Team} (2021).
\newblock {\em R: A Language and Environment for Statistical Computing}.
\newblock R Foundation for Statistical Computing, Vienna, Austria.

\bibitem[Ro et~al., 2015]{ro_outlier_2015}
Ro, K., Zou, C., Wang, Z., and Yin, G. (2015).
\newblock Outlier detection for high-dimensional data.
\newblock {\em Biometrika}, 102(3):589--599.

\bibitem[Roberts et~al., 2015]{roberts_adaptive_2015}
Roberts, S., Martin, M.~A., and Zheng, L. (2015).
\newblock An {Adaptive}, {Automatic} {Multiple}-{Case} {Deletion} {Technique}
  for {Detecting} {Influence} in {Regression}.
\newblock {\em Technometrics}, 57(3):408--417.

\bibitem[Rousseeuw and Leroy, 1987]{Rousseeuw1987}
Rousseeuw, P.~J. and Leroy, A.~M. (1987).
\newblock {\em Robust Regression and Outlier Detection}.
\newblock John Wiley \& Sons, Inc., USA.

\bibitem[Schnabel and Eilers, 2009]{SchnabelEilers2009}
Schnabel, S. and Eilers, P. (2009).
\newblock Optimal expectile smoothing.
\newblock {\em Computational Statistics and Data Analysis}, 53(12):4168--4177.

\bibitem[She and Owen, 2010]{she_outlier_2010}
She, Y. and Owen, A.~B. (2010).
\newblock Outlier {Detection} {Using} {Nonconvex} {Penalized} {Regression}.
\newblock {\em arXiv:1006.2592 [cs, stat]}.
\newblock arXiv: 1006.2592.

\bibitem[Sherif et~al., 2014]{sherif_cbrain_2014}
Sherif, T., Rioux, P., Rousseau, M.-E., Kassis, N., Beck, N., Adalat, R., Das,
  S., Glatard, T., and Evans, A.~C. (2014).
\newblock Cbrain: a web-based, distributed computing platform for collaborative
  neuroimaging research.
\newblock {\em Frontiers in Neuroinformatics}, 8:54.

\bibitem[Smucler and Yohai, 2017]{smuclerRobustSparseEstimators2017}
Smucler, E. and Yohai, V.~J. (2017).
\newblock Robust and sparse estimators for linear regression models.
\newblock {\em Computational Statistics \& Data Analysis}, 111:116--130.

\bibitem[Sobotka and Kneib, 2012]{SobotkaKneib2012}
Sobotka, F. and Kneib, T. (2012).
\newblock Geoadditive expectile regression.
\newblock {\em Computational Statistics and Data Analysis}, 56(4):755--767.

\bibitem[Sobotka et~al., 2014]{expectreg}
Sobotka, F., Schnabel, S., Waltrup, L.~S., Eilers, P., Kneib, T., and
  Kauermann, G. (2014).
\newblock {\em expectreg: Expectile and Quantile Regression}.
\newblock R package version 0.39.

\bibitem[Tibshirani, 1996]{tibshiraniRegressionShrinkageSelection1996}
Tibshirani, R. (1996).
\newblock Regression {Shrinkage} and {Selection} {Via} the {Lasso}.
\newblock {\em Journal of the Royal Statistical Society: Series B
  (Methodological)}, 58(1):267--288.
\newblock \_eprint:
  https://onlinelibrary.wiley.com/doi/pdf/10.1111/j.2517-6161.1996.tb02080.x.

\bibitem[Wang et~al., 2007]{WangLiJiang2007}
Wang, H., Li, G., and Jiang, G. (2007).
\newblock Robust regression shrinkage and consistent variable selection through
  the lad-lasso.
\newblock {\em Journal of Business \& Economic Statistics}, 25(3):347--355.

\bibitem[Wang et~al., 2018a]{wang_multiple_2018}
Wang, T., Li, Q., Chen, B., and Li, Z. (2018a).
\newblock Multiple outliers detection in sparse high-dimensional regression.
\newblock {\em Journal of Statistical Computation and Simulation},
  88(1):89--107.

\bibitem[Wang et~al., 2018b]{wang_multiple_case_2018}
Wang, T., Li, Q., Zang, Q., and Li, Z. (2018b).
\newblock A multiple-case deletion approach for detecting influential points in
  high-dimensional regression.
\newblock {\em Communications in Statistics - Simulation and Computation},
  pages 1--18.

\bibitem[Wang and Li, 2017]{wang_outlier_2017}
Wang, T. and Li, Z. (2017).
\newblock Outlier detection in high-dimensional regression model.
\newblock {\em Communications in Statistics - Theory and Methods},
  46(14):6947--6958.

\bibitem[Zhao et~al., 2013]{zhao_high_dimensional_2013}
Zhao, J., Leng, C., Li, L., and Wang, H. (2013).
\newblock High-dimensional influence measure.
\newblock {\em The Annals of Statistics}, 41(5):2639--2667.

\bibitem[Zhao et~al., 2019]{Zhao_2019}
Zhao, J., Liu, C., Niu, L., and Leng, C. (2019).
\newblock Multiple influential point detection in high dimensional regression
  spaces.
\newblock {\em Journal of the Royal Statistical Society: Series B (Statistical
  Methodology)}, 81(2):385--408.

\end{thebibliography}
\end{document}